\documentclass[aps,prd,twocolumn,nofootinbib,preprintnumbers]{revtex4}

\usepackage{epsf,epsfig,subfigure,graphicx,amsmath,amssymb}
\usepackage{color}

\graphicspath{{Figures/}}

\definecolor{red1}{cmyk}{0,1,1,0.1}
\definecolor{blue1}{cmyk}{1,0,0,0}

\begin{document}

\vskip 0.2in

\title{\bf Boosted Dark Matter at the Deep Underground Neutrino Experiment}

\author{Haider Alhazmi$^{(a,b)}$, Kyoungchul Kong$^{(a, c)}$, Gopolang Mohlabeng$^{(a)}$ and Jong-Chul Park$^{(d)}$}

\affiliation{$^{(a)}$Department of Physics and Astronomy, University of Kansas, Lawrence, KS 66045, USA\\
$^{(b)}$Department of Physics, Jazan University, Jazan 45142, Saudi Arabia\\
$^{(c)}$Pittsburgh Particle physics, Astrophysics, and Cosmology Center, Department of Physics and Astronomy, University of Pittsburgh, Pittsburgh, PA 15260, USA\\
$^{(d)}$Department of Physics, Chungnam National University, Daejeon 34134, Republic of Korea
}

\preprint{PITT-PACC-1615}

\begin{abstract}
We investigate the detection prospects of a non-standard dark sector in the context of boosted dark matter.
We consider a scenario where two stable particles have a large mass difference and the heavier particle 
accounts for most of dark matter in our current universe. 
The heavier candidate is assumed to have no interaction with the standard model particles at tree-level, hence evading existing constraints. Although subdominant, the lighter dark matter particles are efficiently produced via pair-annihilation of the heavier ones in the center of the Galaxy or the Sun.
The large Lorentz boost enables detection of the non-minimal dark sector in large volume terrestrial experiments via exchange of a light dark photon with electrons or nuclei.
Various experiments designed for neutrino physics and proton decay are examined in detail, including Super-K and Hyper-K.
In this study, we focus on the sensitivity of the far detector at the Deep Underground Neutrino Experiment for boosted dark matter produced in the center of the Sun, and compare our findings with recent results for boosted dark matter produced in the galactic center. \\ \\
\end{abstract}


\maketitle

\section{Introduction}

The existence of dark matter (DM) has been firmly established via diverse astrophysical and cosmological observations at multiple scales.
Yet its fundamental nature still remains unidentified.
Since the standard model (SM) of elementary particle physics does not provide a dark matter candidate, new physics must be involved.
Although a variety of searches have been performed to look for dark matter particles, no firm detection has been made thus far and only some tantalizing hints have been reported \cite{Arrenberg:2013rzp}.

Among a plethora of possibilities, models with multiple DM candidates are very well motivated and their phenomenology has been studied at many different scales, from cosmology to the LHC.
Especially on the cosmological side, several issues have been investigated to reconcile a discrepancy between observations and N-body simulations based on cold dark matter (CDM), which include the ``core vs cusp problem''\footnote{Simulations show a steep density profile, while observations of dwarf galaxies indicate a cored density profile~\cite{deBlok:2009sp}.} as well as the ``too big to fail problem''\footnote{Simulations predict that CDM evolves to very dense subhalos of Milky Way type galaxies, which can not host the brightest satellites, but it would be hard to miss the observation of these substructures~\cite{BoylanKolchin:2011de, BoylanKolchin:2011dk}.}.
Warm dark matter has been proposed as a solution to these problems, since it develops a shallower density profile at small scales and therefore avoids unreasonably dense subhalos \cite{Lovell:2013ola}.
Another possible solution to the problems is to introduce self-interaction (SI) between the dark matter particles ($\chi$) \cite{Spergel:1999mh}.
Self-interacting dark matter (SIDM) with $\sigma_{\chi\chi}/m_\chi \sim {\cal O} (1)~ {\rm cm^2/g}$ is known to reconcile the inconsistency between simulations and observations at small scales, while not affecting good CDM behavior at large scales \cite{Rocha:2012jg,Peter:2012jh}.
Here $\sigma_{\chi\chi}$ is the DM self-interaction cross section and $m_\chi$ is the mass of DM.
The matter distribution of the Bullet Cluster \cite{Randall:2007ph} and kinematics of dwarf spheroidals \cite{Zavala:2012us} provide bounds on the size of the self-interaction,
$0.1 ~{\rm cm^2/g} < \sigma_{\chi\chi}/m_\chi < 1.25 ~{\rm cm^2/g}$.

In this paper, we explore the detection prospects of two-component DM at the Deep Underground Neutrino Experiment (DUNE) and Super-K(SK)/Hyper-K(HK).
We focus on a scenario with a large mass gap between the two components,
where the heavier DM has no interaction with SM particles at tree-level, 
while the light one interacts with the heavier counterpart as well as the SM particles.
If the heavier DM is predominant in our current universe,
the dark sector with such candidates is hidden, evading all current direct and indirect detection bounds.
Although the light DM particles are subdominant,
they may be produced via pair-annihilation of the heavier ones even in the current universe, with a large boost due to the large mass difference.
Such boosted DM (BDM) arises in various multi-component DM scenarios \cite{D'Eramo:2010ep, Belanger:2011ww, Belanger:2012vp, DiFranzo:2016uzc} and
recently the discovery potential of BDM in large volume neutrino telescopes has been examined \cite{Huang:2013xfa, Agashe:2014yua, Berger:2014sqa, Kong:2014mia, Kopp:2015bfa}.
In Ref. \cite{Agashe:2014yua}, the heavier DM annihilates in the center of the Galaxy, and its pair-annihilation products travel to the Earth and
leave Cherenkov light in the detector via a neutral current-like interaction, which points toward the galactic center (GC).
The detection of BDM from the Sun (solar BDM) has been studied in Ref. \cite{Berger:2014sqa}, where a search for proton tracks pointing toward the Sun is proposed.
Implication of self-interaction in the context of solar BDM has been discussed in Ref. \cite{Kong:2014mia}.
More recently, Ref. \cite{Necib:2016aez} studied the sensitivity of DUNE and SK, for BDM from the GC and dwarf spheroidal galaxies.

In this study, we investigate the sensitivity of DUNE for BDM coming from the center of the Sun, including self-interaction.
The far detector at DUNE consists of Liquid Argon Time Projection Chambers (LArTPC), which provide excellent particle identification (ID), a good angular resolution and a lower threshold energy \cite{Acciarri:2015uup, Acciarri:2016ooe, Acciarri:2016crz, Strait:2016mof}.
Thanks to the excellent angular resolution and particle ID, we can efficiently reduce expected background events at DUNE, which thus greatly improves the detection prospect of the BDM signal.
We compare our results with those for SK and HK, which are based on Cherenkov radiation.

This paper is organized as follows.
We give a brief review of the model in section \ref{sec:model}, and
a short overview on DUNE and SK/HK in section \ref{sec:exp}.
We study their discovery potential for BDM from the GC in section \ref{sec:gc} and from the Sun in section \ref{sec:sun}, respectively.

\section{Short review of the model\label{sec:model}}
%
\begin{figure*}[t]
\begin{center}
\hspace{0.3cm}
\includegraphics[width=0.3\linewidth]{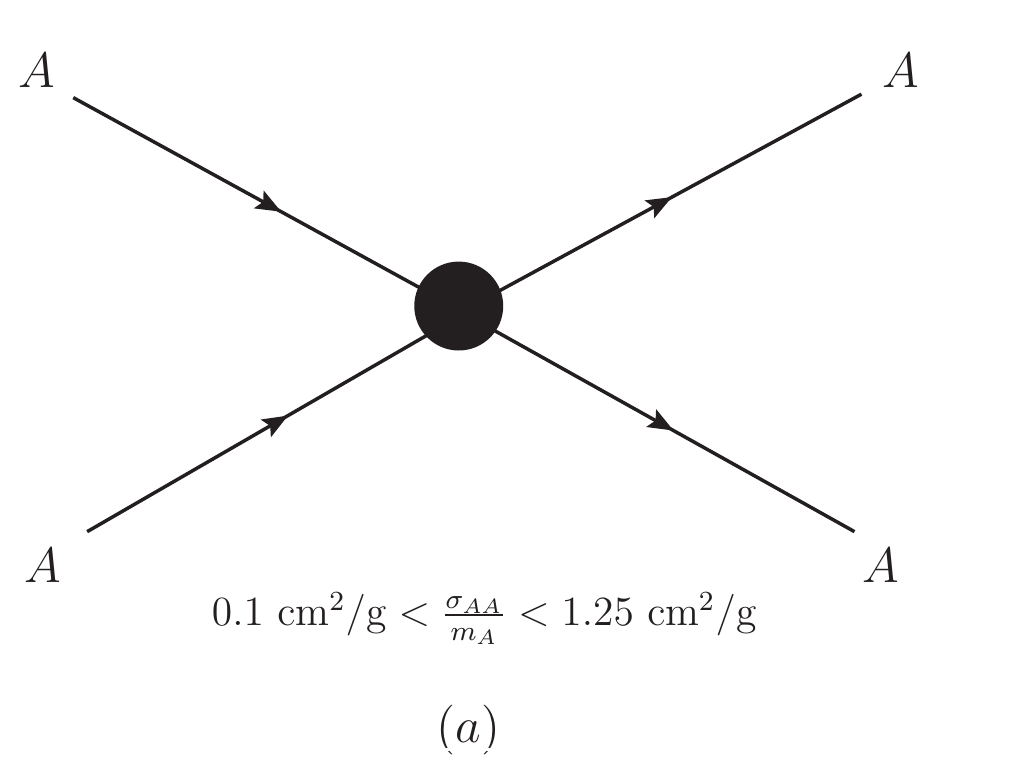}
\hspace*{0.01cm}
\includegraphics[width=0.3\linewidth]{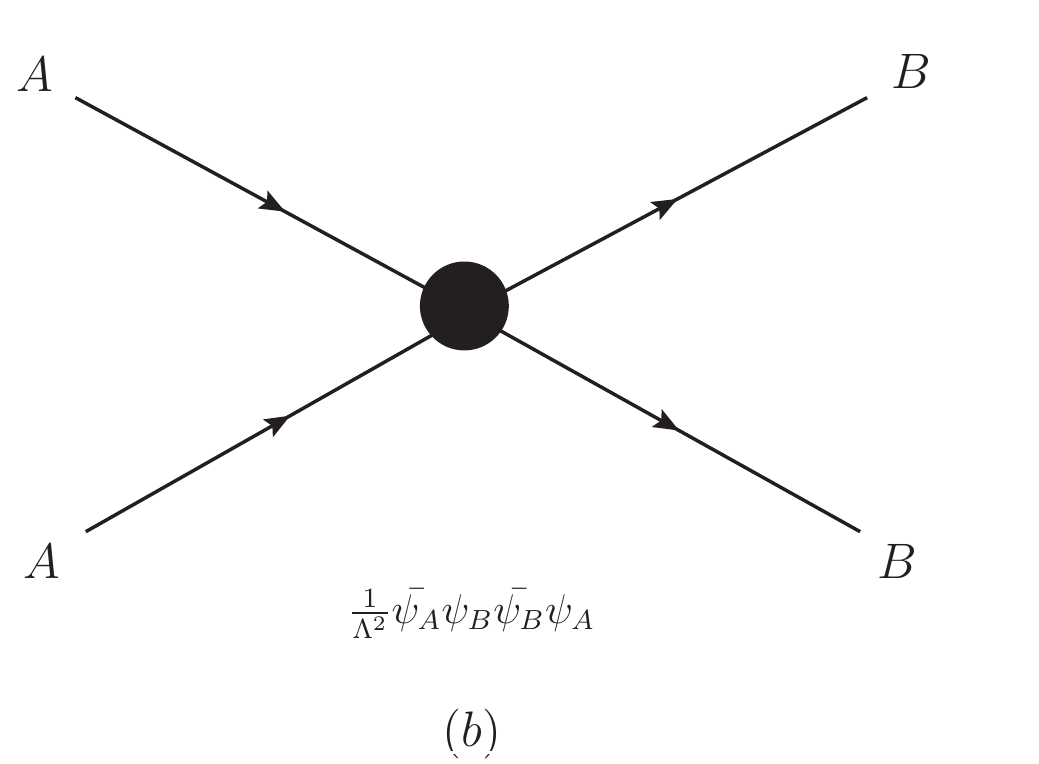}
\hspace*{0.01cm}
\includegraphics[width=0.28\linewidth]{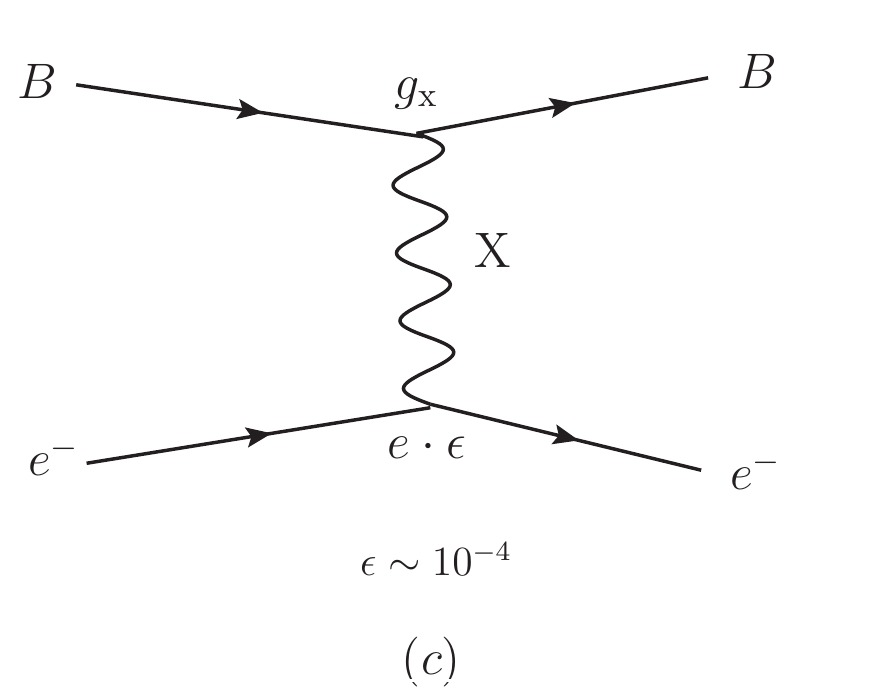}
\end{center}
\vspace*{-0.3cm}
\caption{
Diagrams for (a) self-interaction of the heavier DM component $\psi_A$, (b) production of the boosted DM $\psi_B$ from the annihilation of $\psi_A$, and (c) elastic scattering of $\psi_B$ off an electron.
}
\label{Diagrams}
\end{figure*}
%

We consider a non-minimal DM scenario with two DM species $\psi_A$ and $\psi_B$, whose stability is achieved with two separate symmetries, e.g. ${\rm U(1)}'\otimes{\rm U(1)}''$ or $Z_2\otimes Z_2'$~\cite{Belanger:2011ww, Agashe:2014yua, Kong:2014mia}.
The heavier species $\psi_A$ is the dominant DM component in the universe and has interaction only with $\psi_B$ at tree-level via a contact operator,
\begin{eqnarray}\label{ContactOp}
\mathcal{L}_{AB} = \frac{1}{\Lambda^2} \overline{\psi}_A \psi_B \overline{\psi}_B \psi_A \, ,
\end{eqnarray}
in addition to self-interaction among $\psi_A$ in the range of $0.1\,{\rm cm}^2/{\rm g} < \sigma_{AA}/m_A < 1.25\,{\rm cm}^2/{\rm g}$ (Fig. \ref{Diagrams}(a)), favored by observations and simulations~\cite{Spergel:1999mh, Rocha:2012jg, Randall:2007ph, Zavala:2012us}.
On the other hand, the lighter subdominant DM component $\psi_B$ directly couples to the SM sector.
The lighter DM $\psi_B$ (of mass $m_B$) is produced from the pair-annihilation of the heavier DM $\psi_A$ (of mass $m_A$) in our current universe via the contact interaction Eq.~(\ref{ContactOp}) (Fig. \ref{Diagrams}(b)), and the produced $\psi_B$ is highly boosted due to the large mass difference between $m_A$ and $m_B$.

The lighter species $\psi_B$ is charged under a dark ${U(1)}_X$ gauge symmetry, with a charge $Q_X^B=+1$, which is assumed to be spontaneously broken leading to the mass $m_X$.
In addition, the dark sector is assumed to communicate to the SM sector only through a kinetic mixing between $U(1)_X$ and $U(1)_{\rm EM}$ (originally $U(1)_Y$)~\cite{Okun:1982xi, Holdom:1985ag, Huh:2007zw, Chun:2008by, Chun:2010ve, Park:2012xq, Belanger:2013tla},
\begin{eqnarray}\label{mixing}
\mathcal{L} \supset -\frac{1}{2}\, \sin\epsilon\, X_{\mu\nu}F^{\mu\nu}\,.
\end{eqnarray}
Therefore, $\psi_B$ can elastically scatter off SM particles via a $t-$channel $X$ boson exchange as shown in Fig. \ref{Diagrams}(c).
We will take the dark gauge coupling $g_X$ to be large enough, {\it e.g.,} $g_X=0.5$, so that the large annihilation cross section for $\psi_B \overline{\psi}_B \to XX$ induces a small thermal relic density of $\psi_B$.
We refer to Ref.~\cite{Chun:2010ve} for a dedicated study on the kinetic mixing and hidden sector DM, and Refs.~\cite{Belanger:2011ww, Agashe:2014yua} for computation of exact relic abundances of $\psi_A$ and $\psi_B$.

This model is described by seven parameters:
\begin{equation}
\{m_A, m_B, m_X, \Lambda, g_X, \epsilon, \sigma_{AA}\}\,,
\end{equation}
where $\Lambda$ will be appropriately adjusted to yield the required DM relic abundance, $\Omega_A h^2\simeq \Omega_{\rm DM}h^2 \approx 0.1$, as done in Ref.~\cite{Agashe:2014yua}.
In our analysis, a mass hierarchy of $m_A > m_B > m_X$ is assumed.
$g_X$ and $\epsilon$ always appear as a combination of $(g_X \cdot \epsilon)$ in all the interactions between DM and SM sector particles.
Thus, our study simply depends on five parameters, $\{ m_A, m_B, m_X, g_X \cdot \epsilon, \sigma_{AA} \}$.
For convenient comparison, we take the same benchmark scenario as in Ref.~\cite{Necib:2016aez},
\begin{equation}\label{benchmark}
m_X = 15 ~{\rm MeV},~ g_X = 0.5,~ {\rm and}~ \epsilon^2 = 2\times10^{-7}\,.
\end{equation}
We also choose $\epsilon^2=10^{-8}$ as another benchmark for comparison with Ref.~\cite{Kong:2014mia}.
$\epsilon^2=2\times10^{-7}$ is marginally consistent with current search limits on a hidden $X$ gauge boson (or a dark photon) for $m_X = 15$ MeV~\cite{Batley:2015lha, Ilten:2015hya, Anastasi::2016lwm, Banerjee:2016tad}.

\section{Experimental Details\label{sec:exp}}

The DUNE far detector, which consists of four LArTPC modules to be located deep underground at the Sanford Underground Research Facility, South Dakota, provides an excellent opportunity for particle physics beyond the primary mission of the experiment.
This includes indirect detection of DM using neutrinos, which are produced via the pair-annihilation of DM in the GC or in the center of the Sun.
The excellent angular resolution and particle identification capability of the LArTPC detector would substantially reduce the background in the direction of the expected DM-induced neutrino signal, and could potentially provide competitive limits in the low DM-mass range.
In this paper, we consider direct detection of BDM with the DUNE LArTPC rather than detecting neutrinos induced by DM annihilation.
We compare results against those for other neutrino detectors based on Cherenkov radiation, such as SK and HK.
Table \ref{table:exp} summarizes detector volume, threshold energy and angular resolution for SK, HK and DUNE.
\begin{table}[t]
\begin{center}
\begin{tabular}{|c||c|c|c|c|}
\hline
		&   ~Volume  ~   	&   ~$E_{\rm th}$ ~   &      ~$\theta_{\rm res}$~ &       ~Running Time  ~       \\
		&   ~~(kTon)  ~~   	&   ~~(MeV)~~   		&   ~($^\circ$)~   &        ~ (years)~     \\
\hline   \hline
SK \cite{Fechner:2009aa} 		&           22.5             	&         100                 	 &    3$^\circ$              			&   $>$ 13.6   \\
\hline
HK \cite{Kearns:2013lea}   	&         560               	&        100                  	 &   3$^\circ$					&                                     \\
\hline
~DUNE~\cite{Acciarri:2015uup}~	&    40-50    	&     30  	& 1$^\circ$ 			&              \\
\hline
\end{tabular}
\end{center}
\caption{List of experiments (SK, HK, and DUNE) studied in this paper with volume, threshold energy, angular resolution, and running time.
In principle, the threshold energy at SK/HK could be lowered below 100 MeV at the cost of having worse energy and angular resolution.
However, in our study we use 100 MeV to reduce backgrounds from solar neutrinos and muon decays \cite{Agashe:2014yua}.
We consider two different sizes for DUNE: 10 kTon (DUNE 10) and 40 kTon (DUNE 40), since the staged implementation of the far detector as four 10 kTon modules is planned \cite{Acciarri:2015uup, Acciarri:2016ooe, Acciarri:2016crz, Strait:2016mof}.}
\label{table:exp}
\end{table}

The dominant backgrounds for BDM come from atmospheric neutrinos in the mass range of our interests, while solar neutrinos dominate the background below energies around 20 MeV \cite{Gaisser:2002jj}.
Another background is from muons which do not Cherenkov-radiate but decay to neutrinos in the SK/HK detector.
The relevant energy range for the muon background is about 30--50 MeV and can be alleviated via fiducial volume cuts \cite{Bays:2011si}.
Table \ref{table:exp} shows 100 MeV for threshold energy at SK/HK.
However, in principle the threshold energy at SK can be lowered even below 10 MeV.
For example, Ref. \cite{Abe:2010hy} studies solar neutrinos, focusing on the 5--20 MeV range.
In this case, both energy resolution and angular resolution become poor,
$\frac{\sigma(E)}{E} > 0.15$ and $\theta_{\rm res} > 25^\circ$ for $E_e < 10$ MeV \cite{Abe:2010hy}.
We use $E_e > 100$ MeV in our analysis, to reject backgrounds from solar neutrino and muon decays.
However, this cut may be lowered down to 50--100 MeV with slightly poorer angular resolution \cite{Abe:2010hy, Agashe:2014yua}.
For DUNE, the muon background can be distinguishable due to excellent particle ID and we use $E_{\rm th}=30$ MeV as described in the DUNE CDR \cite{Acciarri:2015uup, Acciarri:2016ooe, Acciarri:2016crz, Strait:2016mof}.

For angular resolution of SK/HK, we use $\theta_{\rm res}=3^\circ$ following Ref. \cite{Agashe:2014yua} but in the energy range of our interests $E_e > 100$ MeV, it can be brought down to a lower value.
The angular resolution of SK (single-ring e-like events) is 3$^\circ$ for sub-GeV ($<$ 1.33 GeV) and 1.2$^\circ$ for multi-GeV ($>$ 1.33 GeV) \cite{Pik:2012qsy}.

For backgrounds at SK, we use the fully contained single-ring e-like events including both sub-GeV (0-decay electron events only) and multi-GeV as a conservative estimation for a total of 10.7 years \cite{Dziomba:2012paz} (${N^{\rm all~sky}_{\rm SK}}/{\Delta T} \simeq 923 ~{\rm year^{-1}}$ for 22.5 kTon).
For our discussion in the rest of this paper, we normalize the rate to 13.6 years, which is the current exposure time at SK.\footnote{More data has been used in Ref. \cite{Richard:2015aua} but it does not discriminate 0- and 1-decay electron events.}

\begin{table*}[t]
\begin{center}
\begin{tabular}{|c||c|c|c|c|}
\hline
	& ~~DUNE 10~~ & ~~DUNE 40 ~~ & ~~~SK~~~ & ~~~HK~~~ \\
\hline \hline	
~~~GC~~~  & 1 with $10^\circ$ & 4 with $10^\circ$ & 7.01 with $10^\circ$ & 174 with $10^\circ$ \\
\hline
~~~Sun~~~ & ~~0.01 with $1^\circ$ ~~&  ~~0.04 with $1^\circ$~~ 	&  ~~0.632 with $3^\circ$~~	 &  ~~15.7 with $3^\circ$ ~~ \\
\hline
\end{tabular}
\end{center}
\caption{Expected number of background events per year with appropriate angular cut and threshold energy.}
\label{table:bknd}
\end{table*}

In the case of BDM from the GC, the number of expected signal events is obtained within a cone of half angle $\theta_C \simeq 10^\circ$ for maximum sensitivity, and the backgrounds are calculated correspondingly, $\frac{N^{\theta_C}_{\rm SK}}{\Delta T} = \frac{1 - \cos\theta_C}{2} \frac{N^{\rm all~sky}_{\rm SK}}{\Delta T} \simeq 7.01 ~{\rm year^{-1}}$ for 22.5 kTon \cite{Agashe:2014yua}.
LArTPC detectors have several advantages over Cherenkov-based detectors, such as lower threshold energy, better angular resolution and efficient vetoing of events with hadronic activities \cite{Necib:2016aez}.
These features are useful in identifying BDM signals and reduce the number of background events.
A background-study at DUNE 10 (DUNE with 10 kTon) by simulation using the GENIE neutrino Monte-Carlo software results in a conservative estimate of background events,
${N^{\rm all~sky}_{\rm DUNE 10}}/{\Delta T} \simeq 128 ~{\rm year^{-1}}$ for 10 kTon \cite{Necib:2016aez},\footnote{According to DUNE CDR~\cite{Acciarri:2015uup}, the expected number of fully contained electron-like events including oscillations is 14053/(350 kTon$\cdot$year), which corresponds to $402 ~{\rm year^{-1}}$ for DUNE 10.
In Ref.~\cite{Necib:2016aez}, they however find that less than 32\% of background events pass the hadronic cuts which are implemented after simulation.
As a result, for DUNE 10, $402 ~{\rm year^{-1}} \times 32\% \simeq 128 ~{\rm year^{-1}}$ is obtained.} and thus for the GC,
$\frac{N^{\theta_C}_{\rm DUNE 10}}{\Delta T} = \frac{1 - \cos\theta_C}{2} \frac{N^{\rm all~sky}_{\rm DUNE 10}}{\Delta T} \simeq 1 ~{\rm year^{-1}}$.
Background rates for DUNE 40 (DUNE with 40 kTon) and HK are obtained by a simple rescaling based on their volume.
See Table \ref{table:bknd} for a summary of background events used in our study.

In the case of BDM arising from the Sun, angular resolution becomes crucial. 
The number of background events within a cone of angle $\theta$ is proportional to $\frac{1-\cos \theta}{2} \approx \theta^2/4$ for $\theta \ll 1$, and decreases rapidly as $\theta$ decreases.
On the other hand, the number of signal events does not change, as the Sun is effectively a point-like source.
Therefore, $\theta_C$ can be reduced to $\theta_{\rm res}$ as shown in Table \ref{table:exp}, which will reduce the number of background events significantly, while the number of signal events is not affected.
In comparison between SK and DUNE, the angular resolution for SK is $\theta^{\rm SK}_{\rm res} = 3^\circ$ while it is $\theta^{\rm DUNE}_{\rm res} = 1^\circ$ for DUNE.
This implies that background rejection at DUNE would be nine times better than at SK, {\it if all other conditions are identical}.
A change in angular cut from $10^\circ$ for the GC to $1^\circ$ for the Sun reduces background events by a factor of 100 for the same detector.

Another strength of the DUNE detector is a lower threshold energy, $E_{\rm th}=30$ MeV.
This is partly due to excellent particle ID with LArTPC, which also allows better background rejection, {\it i.e.,} rejection of Michel electrons from muon decays \cite{Necib:2016aez}.

The main advantage of SK over DUNE is that it has already been running for more than 13 years and will accumulate more data over the next few years at least.
In addition, its volume is about twice as large as that at DUNE 10, while HK might be 10 times (or more) bigger than DUNE 40.
The phenomenology of BDM with the HK detectors at two different location is also worth investigating \cite{T2HKK}.

\section{Boosted Dark Matter from the Galactic Center\label{sec:gc}}

In this section, we discuss the sensitivity of SK, HK and DUNE on the boosted dark matter arising from the galactic center.

\subsection{BDM flux and signal}

Following the formalism in Ref. \cite{Agashe:2014yua}, we calculate the flux of boosted DM $\psi_{B}$ coming from the galactic center through the annihilation $\psi_A \overline{\psi}_A \to \psi_B \overline{\psi}_B$ as
\begin{align}
\label{eq:fullflux}
\frac{d\Phi_{\rm GC}}{d\Omega dE_{B}} = \frac{r_{\rm Sun}}{16\pi} \left(\frac{\rho_{0}}{m_{A}}\right)^{2} \langle \sigma_{A \bar{A} \rightarrow B \bar{B}} v \rangle\, J\, \frac{dN_{B}}{dE_{B}} \,,
\end{align}
where $r_{\rm Sun} = 8.33$ kpc is the distance from the GC to the Sun, $\rho_{0}$ is the local dark matter density with a value of 0.3 $\rm GeV/cm^3$ and $\langle \sigma_{A \bar{A} \rightarrow B \bar{B}} v \rangle$ is the thermally averaged annihilation cross-section of $\psi_{A}$ into $\psi_{B}$ around the GC.
The galactic halo information is encoded in the so-called $J$-factor which involves an integral over the DM density squared along the line of sight (l.o.s):
\begin{align}
J(\theta) = \int_{\rm l.o.s} \frac{ds}{r_{\rm Sun}} \left(\frac{\rho(r(s, \theta))}{\rho_{0}}\right)^{2}\,.
\end{align}
Here $\rho(r(s, \theta))$ is the galactic halo DM density profile and $s$ is the l.o.s distance from the source to the Earth, while $r(s, \theta) = \sqrt{r_{\rm Sun}^{2} + s^{2} - 2 r_{\rm Sun} \cdot s \cdot  \cos\theta}$ is a coordinate distance centered on the GC and $\theta$ is the angle between the direction of the l.o.s and the GC--Earth axis.
We assume the NFW halo profile~\cite{Navarro:1995iw, Navarro:1996gj} following Ref.~\cite{Agashe:2014yua}.
For the purposes of this study, it is in fact more robust to consider a DM halo profile incorporating SIDM, which ensures the correct DM density at the GC and around our solar system.
Recent studies on SIDM however suggest that for the self-interaction strengths provided by the limits from the Bullet cluster and dwarf spheroidals, density profiles of SIDM are intimately tied with the details of the disk and bulge formation as well as the associated feed back of a baryon dominated galaxy such as the Milky Way.
SIDM profile turns out to be comparable to the NFW profile in our region of interest \cite{Kaplinghat:2015gha, Kaplinghat:2013xca, Kaplinghat:2015aga}.
We assume that the $\psi_{B}$ particles from this process are mono-energetic and thus their differential energy spectrum is simply described by
\begin{eqnarray}\label{B-spectrum}
\frac{dN_B}{dE_B} = 2 \delta(E_B - m_A)\,,
\end{eqnarray}
where $E_{B}$ is the energy of the boosted particle $\psi_{B}$.
Finally, the boosted $\psi_B$ flux over a cone of a half angle $10^\circ$ around the GC can be approximated by~\cite{Necib:2016aez}
\begin{eqnarray}\label{flux-GC-10}
\Phi_{\rm GC}^{10^\circ} &\simeq& 4.7\times10^{-8} {\rm cm}^{-2}{\rm s}^{-1}\, \nonumber\\
& & \times \left( \frac{\langle \sigma_{A\overline{A} \to B\overline{B}} v \rangle}{3\times10^{-26}\, {\rm cm}^3/{\rm s}} \right)\,
\left( \frac{20\, {\rm GeV}}{m_A} \right)^2 \,.
\end{eqnarray}

To mitigate backgrounds, we require the BDM events to fall within a $\theta_{C}$ cone around the GC.
The optimal choice of $\theta_{C}$ is about $10^\circ$ for the annihilation case as discussed in Ref.~\cite{Agashe:2014yua}, which is also used in our analysis.
For BDM interacting with electrons, the number of signal events is given by
\begin{align}
N_{\rm sig}^{\rm GC} = \Delta T ~N_{\rm target} ~\Phi_{\rm GC}^{\theta_{C}} ~\sigma_{Be^{-} \rightarrow Be^{-}} \,,
\label{eqn:signal}
\end{align}
where $\Delta T$ is the exposure time of the experiment and $N_{\rm target}$ is the total number of target electrons in a given experiment, which is proportional to the volume of the experiment.
The quantity $\Phi_{\rm GC}^{\theta_{C}}$ is the flux of BDM particles coming from a $\theta_{C}$ cone around the GC and $\sigma_{Be^{-} \rightarrow Be^{-}}$ is the elastic scattering cross-section between the boosted $\psi_{B}$ and an electron in the experiment.
We refer to Ref. \cite{Agashe:2014yua} for more details.

To compute the sensitivity of each detector for BDM coming from the GC, we use the number of background events listed in Table \ref{table:bknd}.
The signal significance is defined as
\begin{align}
\sigma = \sqrt{   2 \big (N_{\rm sig} + N_{\rm BG} \big ) \log \Big ( 1 + \frac{N_{\rm sig}}{N_{\rm BG}}\Big )- 2 N_{\rm sig}}\, ,
\label{eq:sigma}
\end{align}
where $N_{\rm sig}$ is the number of signal events given by Eq.~(\ref{eqn:signal}) and $N_{\rm BG}$ is the number of background events in a $\theta_{C}$ cone, given in section \ref{sec:exp}.
We have verified that the same results are obtained with a likelihood ratio, assuming a Poisson distribution as in Ref. \cite{Kim:2016plm}.

In order to effectively study the dependence of the signal sensitivity on the threshold energy of the experiment, we reduce the cross-section $\sigma_{Be^{-} \rightarrow Be^{-}}$ to a constant cross-section having assumed a constant scattering amplitude as discussed in Ref. \cite{Necib:2016aez}.
In this limit, we redefine Eq.~(\ref{eqn:signal}) as
\begin{align}
N_{\rm sig}^{\rm GC} = \Delta T ~N_{\rm target} ~\Phi_{\rm GC}^{\theta_{C}} ~\sigma_{0} ~\left(1-\frac{E_{\rm th}}{E_{\rm max}}\right) \,,
\end{align}
where $\sigma_{0}$ is the constant cross-section for $\sigma_{Be^{-} \rightarrow Be^{-}}$. The number of signal events has been rescaled in terms of the threshold energy of the experiment and the maximum energy imparted to an electron after scattering which is given by
\begin{align}
E_{\rm max} = m_{e}\, \frac{(E_{B} + m_{e})^2 + E_{B}^{2} - m_{B}^{2}}{(E_{B} + m_{e})^2 - E_{B}^{2} + m_{B}^{2}} \,,
\end{align}
where $m_{e}$ is the electron mass.
To get the sensitivity in this limit, we use Eq.~(\ref{eq:sigma}) with the same background rates.

\subsection{Detection prospect}

We first reproduced all the results on BDM from the GC in Ref. \cite{Necib:2016aez}, where the performance of SK, HK and DUNE detectors are compared, assuming the same 13.6 years of physics running for all detectors.\footnote{We thank Lina Necib for help and clarification when reproducing results in Ref. \cite{Necib:2016aez}.}
The authors of Ref. \cite{Necib:2016aez} have shown the excellent performance of DUNE with 10 kTon, which is comparable to SK (with twice larger volume).
Moreover, DUNE covers slightly larger parameter space due to the lower threshold energy.

%
\begin{figure*}[t]
\begin{center}
\hspace*{-0.8cm}
\includegraphics[width=0.4\linewidth]{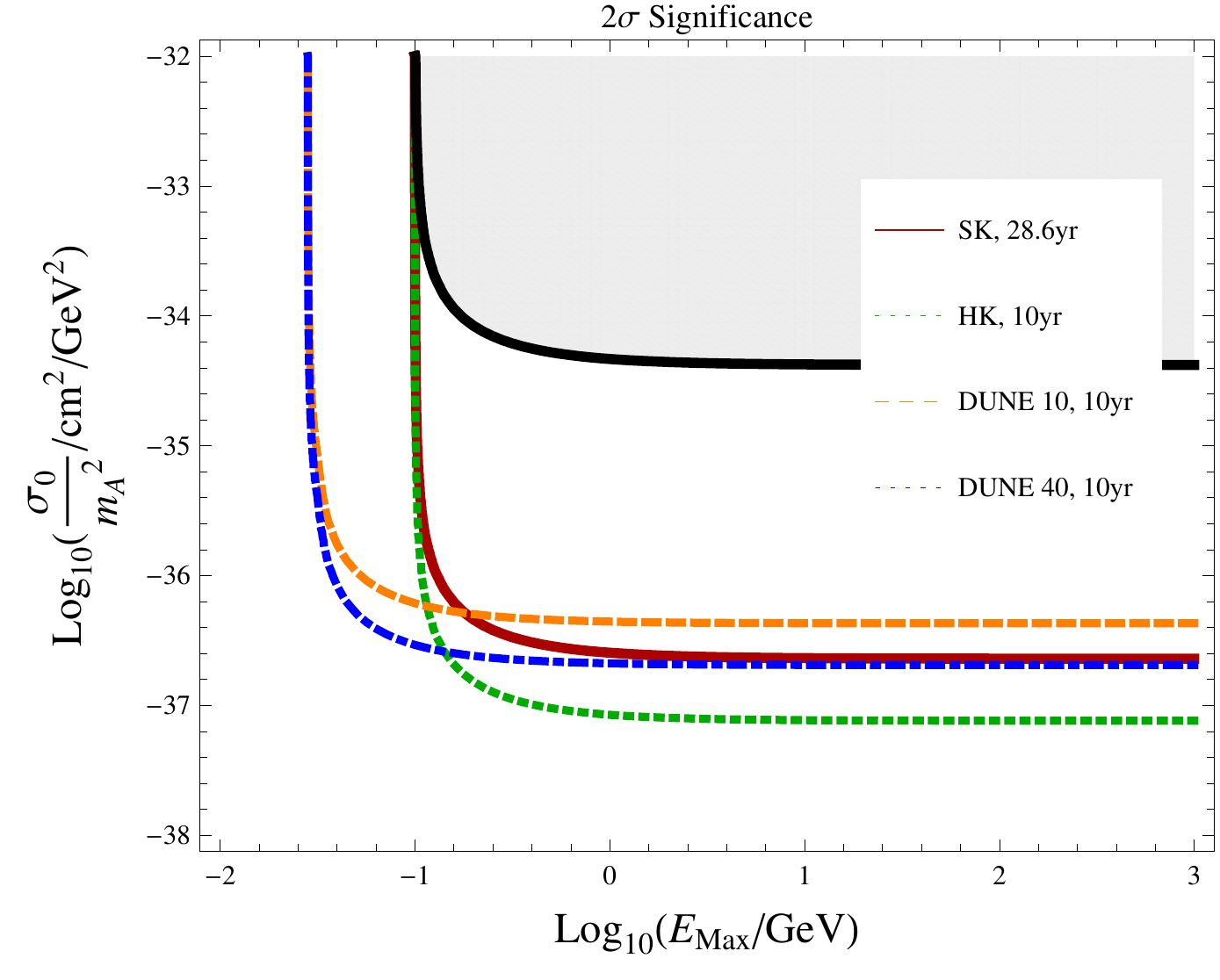}
\hspace*{0.5cm}
\includegraphics[width=0.4\linewidth]{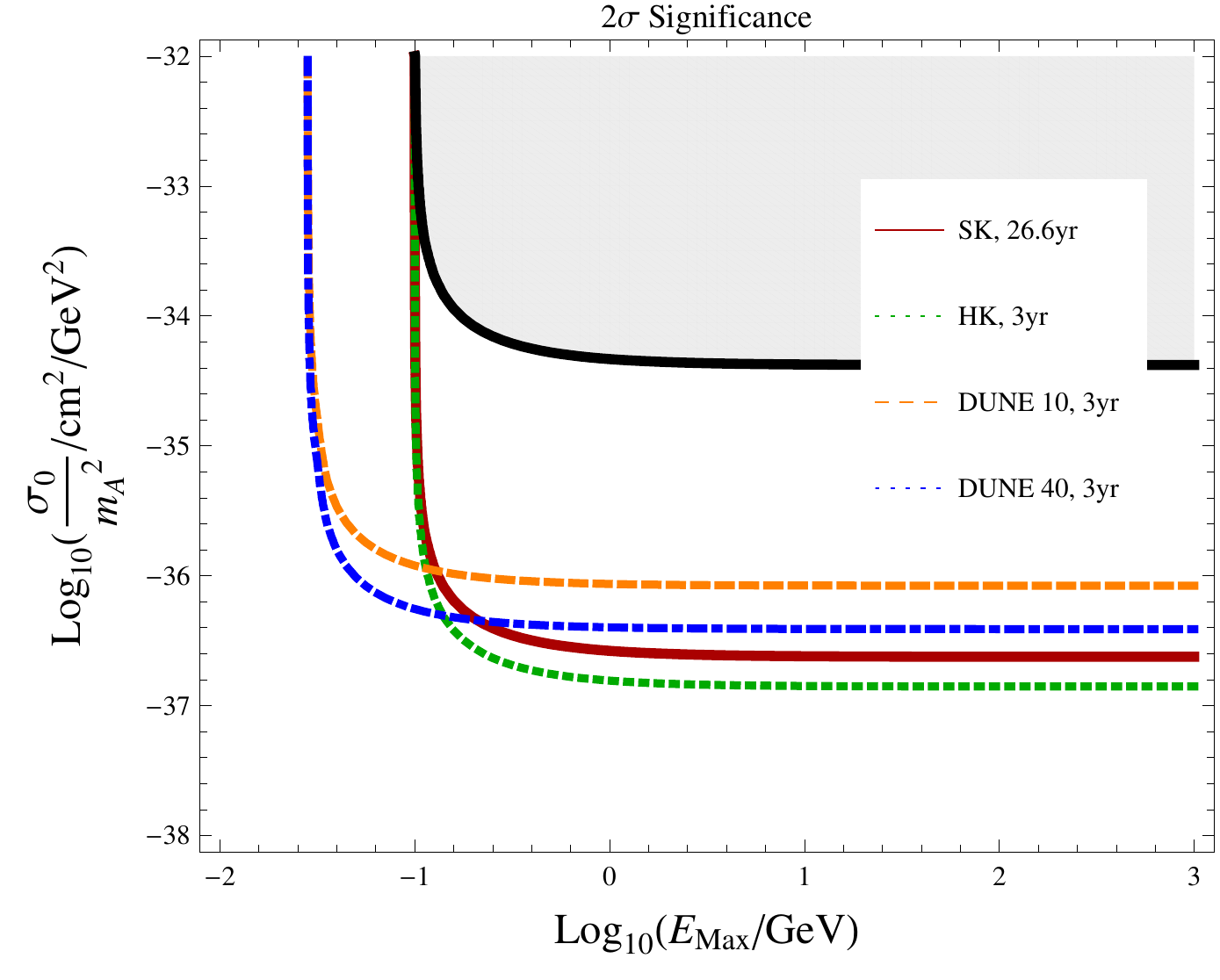} \\
\vspace*{0.2cm}
\includegraphics[width=0.41\linewidth]{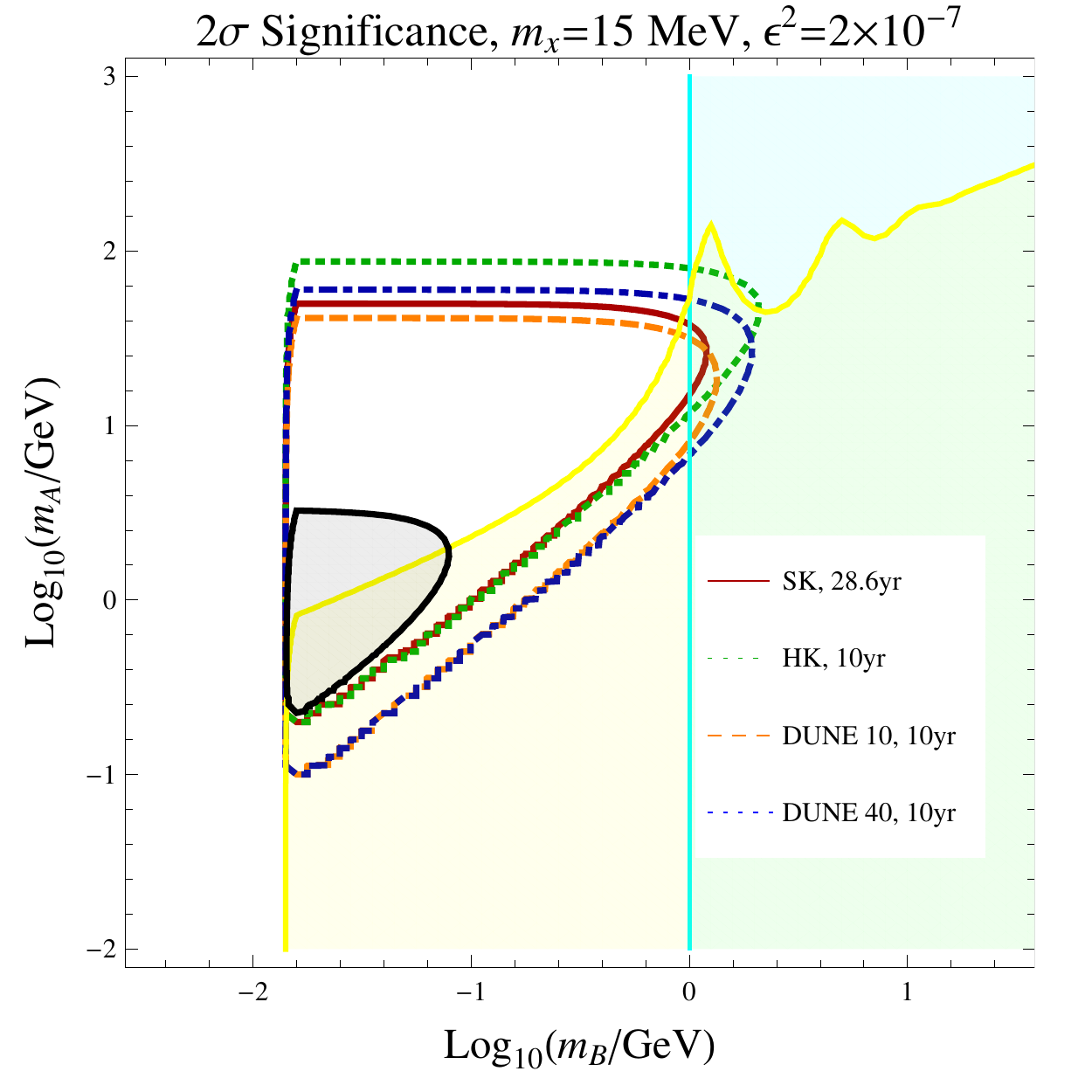}
\hspace*{0.5cm}
\includegraphics[width=0.41\linewidth]{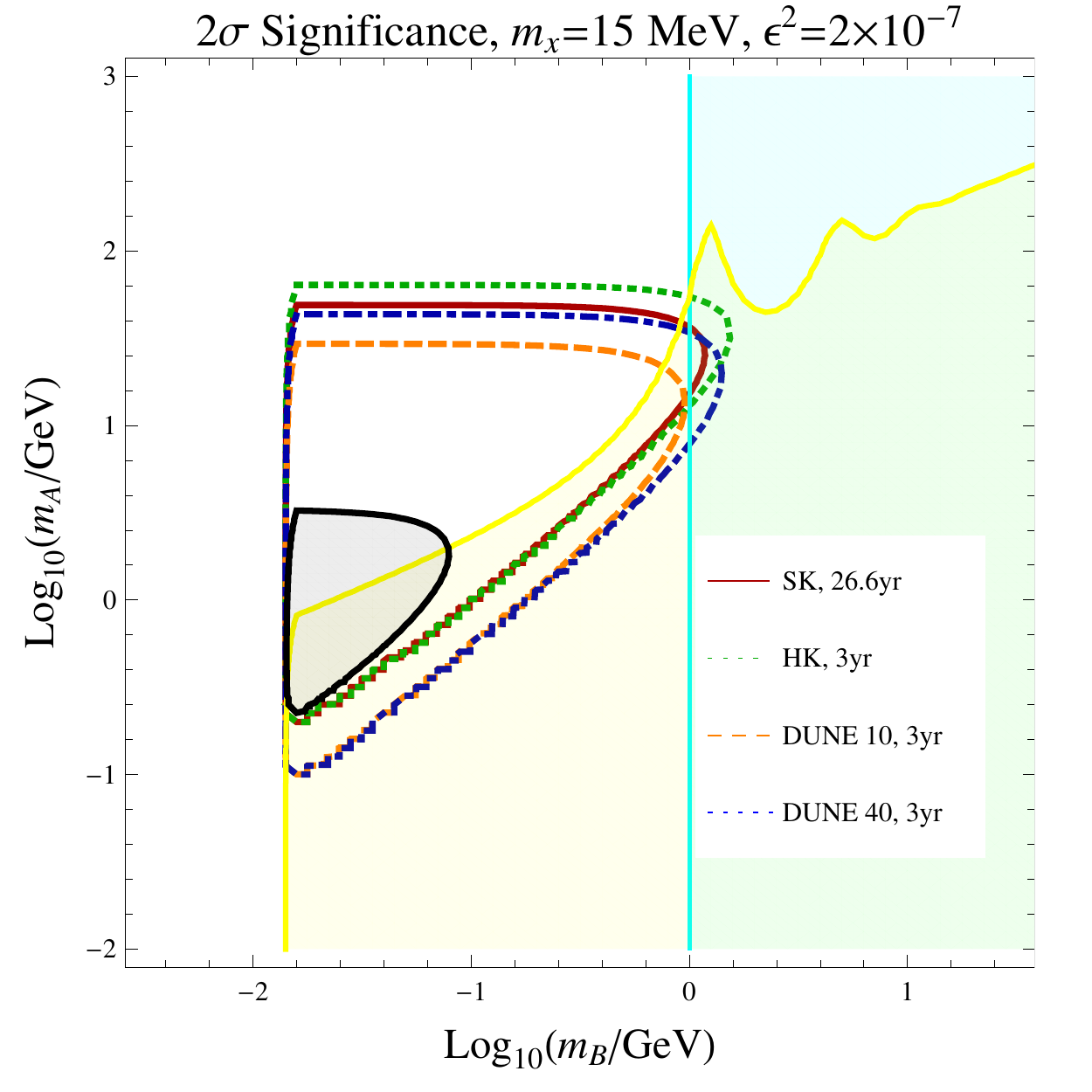}
\end{center}
\vspace*{-.3cm}
\caption{95\% limits on the effective cross section $\sigma_0$ (top panel) and the 2$\sigma$ signal-significance (bottom panel) assuming 5 years of construction and 10 years of physics running of DUNE (left), and 10 years of construction and 3 years of physics running of DUNE (right).
The corresponding total running time of SK would be 28.6 and 26.6 years, respectively.
We assume that the HK timeline is the same as DUNE.
The gray-shade represents the current 2$\sigma$ exclusion with all-sky data from SK, assuming 10\% systematic uncertainty in the background estimation.
The other shaded areas are potential bounds from direct detection of non-relativistic $\psi_B$ (in cyan, with vertical boundary) and CMB constraints on $\psi_B$ annihilation (in yellow, with diagonal boundary).
}
\label{fig:GC}
\end{figure*}

In Fig. \ref{fig:GC}, we show the 2$\sigma$ signal-significance in the $\sigma/m_A^2$--$E_{\rm max}$ plane (top) and in the $m_A$--$m_B$ plane (bottom) for various detectors including SK, HK and DUNE with two different detector sizes (10 kTon and 40 kTon).
We consider two different timelines: 5 years of construction and 10 years of physics running of DUNE in the left panel, and 10 years of construction and 3 years of physics running of DUNE in the right panel.
The total physics running time of SK would be 28.6 and 26.6 years, respectively.
We assume that the HK timeline is the same as the DUNE timeline.
An approximate expression for the flux as in Eq. (\ref{flux-GC-10}) is used only for the two figures in the upper panel of Fig. \ref{fig:GC} (following Ref.~\cite{Necib:2016aez}), while the full flux in Eq. (\ref{eq:fullflux}) is used in all other figures.
We used the corresponding volume, angular resolution and threshold energy for each detector as summarized in Table \ref{table:exp}.\footnote{In Ref. \cite{Necib:2016aez}, $\theta_{res} =5^\circ$ is used.
We checked however that there is no significant difference between the results from 3$^\circ$ and 5$^\circ$ of angular resolutions for the GC analysis.}
We find that the increment in the number of events with $E_{\rm th}=$30 MeV to the number of events with $E_{\rm th}=$100 MeV is about 20--50 \% in the bulk of parameter space of the $m_A$--$m_B$ plane, and the signal increases very rapidly closer to the diagonal direction, $m_A \sim m_B$.

Fig. \ref{fig:GC} also includes the $2\sigma$ exclusion (in gray) using currently available all-sky SK data assuming 10\% systematic uncertainty in the background estimation.
There are other relevant but model-dependent bounds such as the direct detection of non-relativistic $\psi_B$ and CMB constraints \cite{Agashe:2014yua}.
Although the relic abundance of $\psi_B$ is small, it has a large $\psi_B$-nucleon scattering cross section.
The mass range of $\psi_B$ that we are interested in is $m_B \lesssim {\cal O} (1)$ GeV, and the corresponding recoil energy is close to the threshold energy of many direct detection experiments.
The most stringent bounds come from DAMIC \cite{deMelloNeto:2015mca} due to its low threshold energy.
The expected elastic scattering cross section is so large that any events above the threshold energy would be seen, even when taking into account an effective nuclear cross section that is properly scaled down by the non-relativistic relic abundance,
$\sigma^{eff}_{Bp \to Bp} = \frac{\Omega_B}{\Omega_{DM}} \sigma_{Bp \to Bp}$ \cite{Agashe:2014yua}.
From Ref. \cite{deMelloNeto:2015mca}, we conclude that $m_B \ge 1$ GeV is disfavored by DAMIC data if $\psi_B$ couples to quarks, which is shown as the cyan shaded region.
Although sub-GeV DM is better constrained by scattering off electrons than off nuclei \cite{Essig:2011nj}, as in XENON10 bounds \cite{Essig:2012yx}, BDM signals are not affected by XENON10 due to different kinematics, and it turns out that bounds from CMB heating are more important \cite{Agashe:2014yua}, which is shown as the yellow-shaded area.
Other constraints such as limits on the dark photon, direct detection of non-relativistic $\psi_A$, indirect detection of non-relativistic $\psi_B$ and BBN bounds on $\psi_B$ annihilation are either weaker than the CMB bound or evaded by our choice of parameters.
We note that apart from the current SK bound, all other limits are model-dependent and it is certainly possible to avoid or weaken the bounds.
For instance, DM models with $p-$wave annihilation can easily avoid the CMB constraint due to $v^2$ suppression with $v \sim 10^{-3}$, and direct detection bounds do not apply if the non-relativistic $\psi_B$ does not couple to quarks.

\section{Boosted Dark Matter from the Center of the Sun\label{sec:sun}}

In this section, we discuss the sensitivity of SK, HK and DUNE on the boosted dark matter arising from the center of the Sun.

\subsection{BDM flux and DM evolution in the Sun}

We follow the formalism in Ref. \cite{Kong:2014mia} for the calculation of the flux of BDM particles from the Sun.
The BDM flux via pair-annihilation of $\psi_A$ in the Sun ($\psi_A\overline{\psi}_A \to \psi_B\overline{\psi}_B$) is defined as
\begin{align}
\frac{d \Phi_B^{\rm Sun}}{dE_B} = \frac{\Gamma_A^{\psi_A}}{4\pi R_{\rm Sun}^{2}}\, \frac{d N_B}{d E_B} \,.
\label{eq:fluxsun}
\end{align}
Here there is no angular contribution in the differential spectrum since there is no l.o.s integration between the Sun and the Earth, given that the Sun is a point-like source.
The quantity $d N_B/d E_B$ is the differential energy spectrum of BDM $\psi_B$ at the production source assuming two mono-energetic boosted particles, which is again given by Eq.~(\ref{B-spectrum}), and $R_{\rm Sun}$ is the distance from the Sun to the Earth.
$\Gamma_A^{\psi_A}$ is the annihilation rate of relic $\psi_{A}$ particles that are captured inside the Sun and the current rate is given by
\begin{eqnarray}
\Gamma_A^{\psi_A} = \frac{C_a}{2} N_{\psi_A}^2 (t_\odot)\,,
\end{eqnarray}
where $N_{\psi_A}$ is the number of $\psi_{A}$ particles that are captured inside the Sun and $t_\odot \simeq 4.6 \times 10^9$ year is the age of the Sun.
In general, the time evolution of the DM number $N_\chi$ in the Sun is described by the following simple differential equation~\cite{Chen:2014oaa}
\begin{eqnarray}\label{N-evolution}
\frac{dN_\chi}{dt} = C_c + (C_s-C_e)N_\chi - (C_a+C_{se})N_\chi^2\, .
\end{eqnarray}
Here $C_{se}$ is the evaporation rate due to DM self-interaction,
$C_{a}$ the DM annihilation rate,
$C_{e}$ the DM evaporation rate due to DM-nuclei interactions,
$C_s$ the DM self-capture rate, and
$C_c$ is the DM capture rate by the Sun.
All these coefficients have been well-studied and reasonable parameterizations already exist in the literature (see Ref.~\cite{Kong:2014mia} and references therein.).

The BDM particles produced in the center of the Sun may lose their kinetic energy as they travel through the Sun.
This can occur because of the relatively large scattering cross-section with electrons in the Sun.
There can be significant scattering with nucleons; however, the scattering rates are relatively smaller than those with electrons.
We refer to Ref. \cite{Kong:2014mia} for detailed description on DM number evolution, annihilation rate and energy loss inside the Sun, and the final BDM flux.

\subsection{Detection prospect}

We compute the number of signal events for the boosted DM coming from the Sun as
\begin{align}\label{NsigSun}
N_{\rm sig}^{\rm Sun} = \Delta T ~N_{\rm target} ~\Phi_{B}^{\rm Sun} ~\sigma_{Be^{-} \rightarrow Be^{-}} \, ,
\end{align}
where $\Phi_{B}^{\rm Sun}$ is the flux of BDM $\psi_B$ from the Sun.
It is dependent on the size of the self-interaction of relic $\psi_{A}$ in the Sun.
The larger the self-interaction strength,
the more $\psi_{A}$ particles are captured in the Sun, which results in a larger flux of boosted particles.
Just as in the GC case, $\sigma_{Be^{-} \rightarrow Be^{-}}$ is the elastic scattering
cross-section between BDM and electrons.
Again as before, $\Delta T$  and $N_{\rm target}$ are the observation time and the number of target electrons, respectively.
For consistency check, we have reproduced all the results shown in Ref. \cite{Kong:2014mia}.

\begin{figure*}[t]
\begin{center}
\hspace*{0.1cm}
\includegraphics[width=0.4\linewidth]{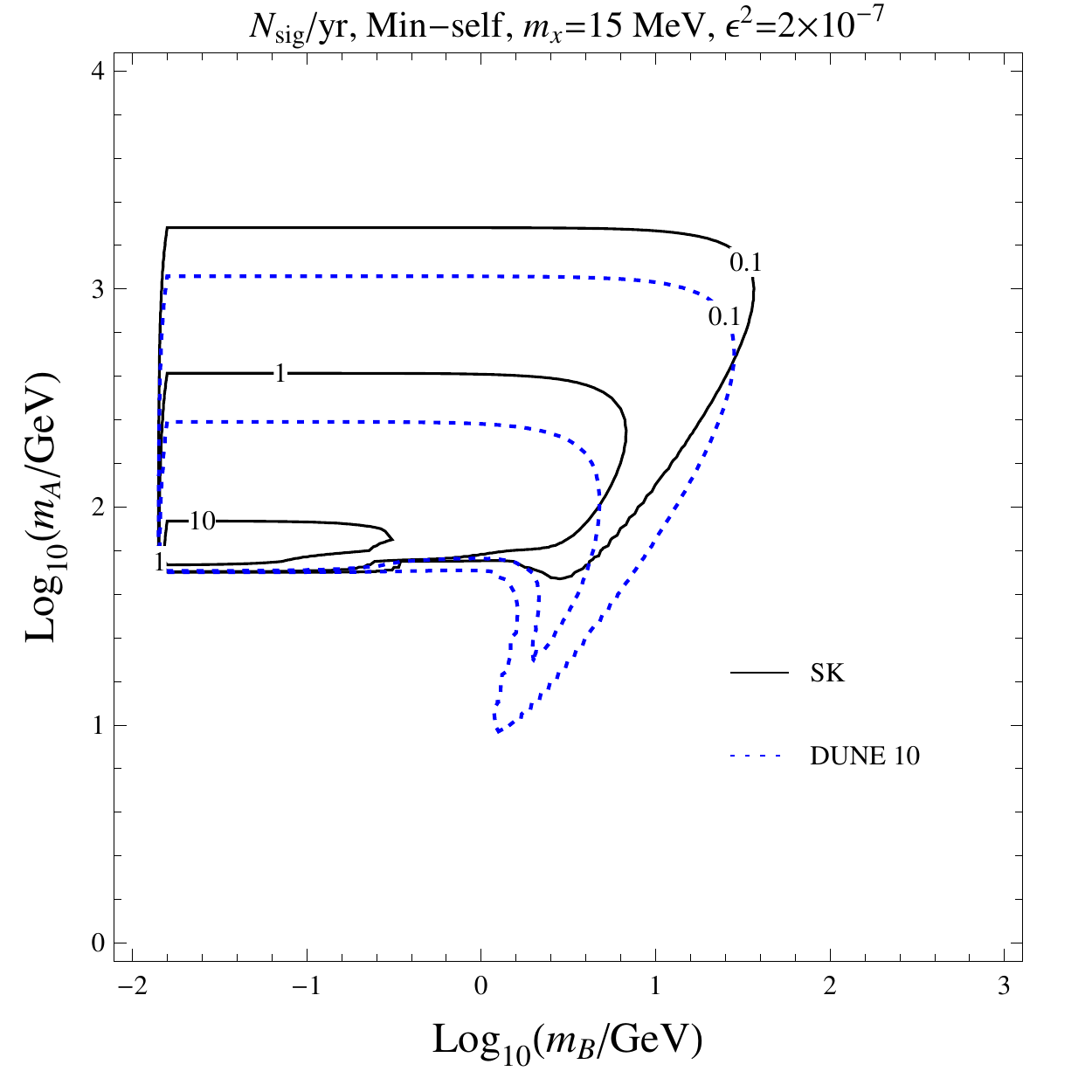}
\hspace*{0.5cm}
\includegraphics[width=0.4\linewidth]{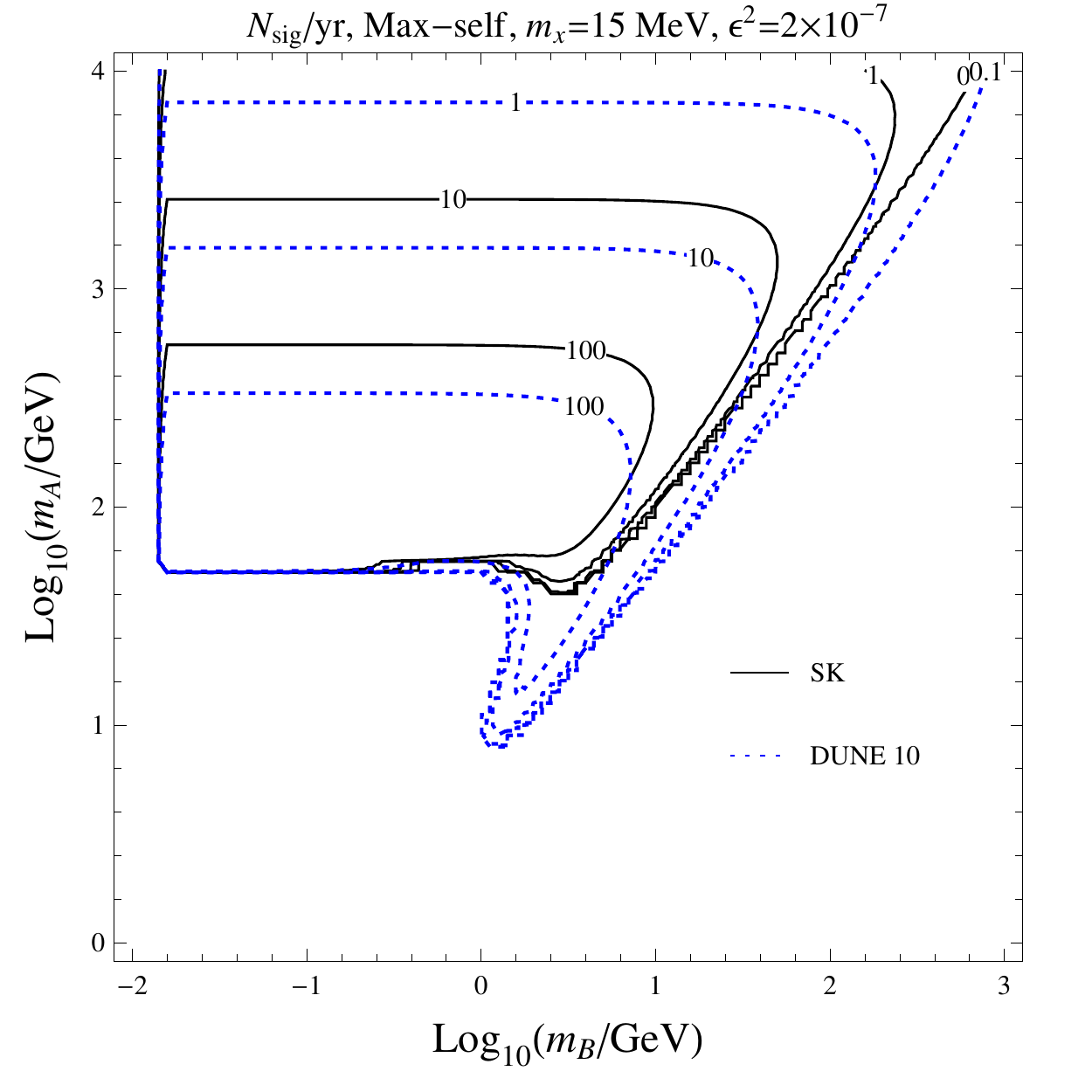} \\
\vspace*{0.2cm}
\includegraphics[width=0.4\linewidth]{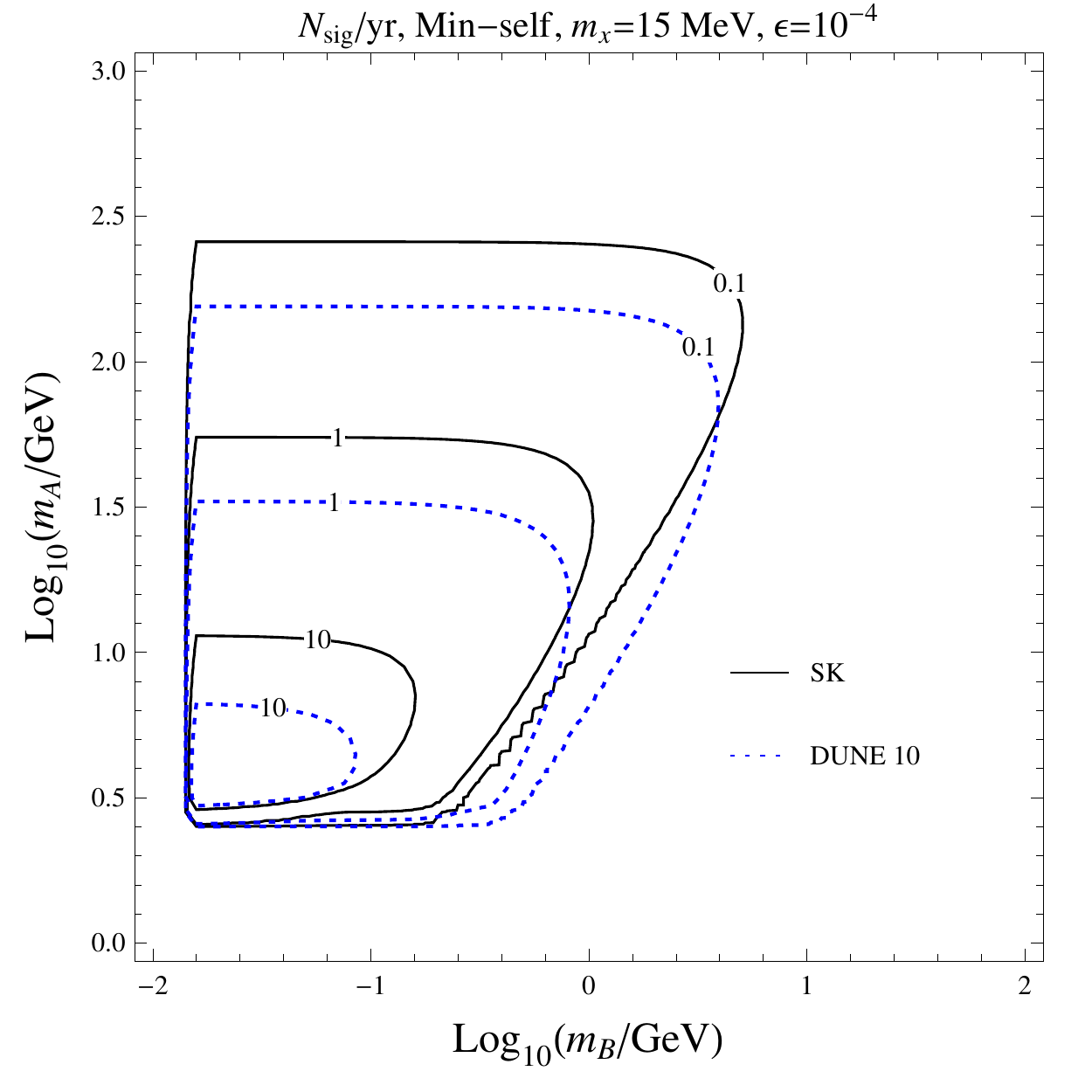}
\hspace*{0.5cm}
\includegraphics[width=0.4\linewidth]{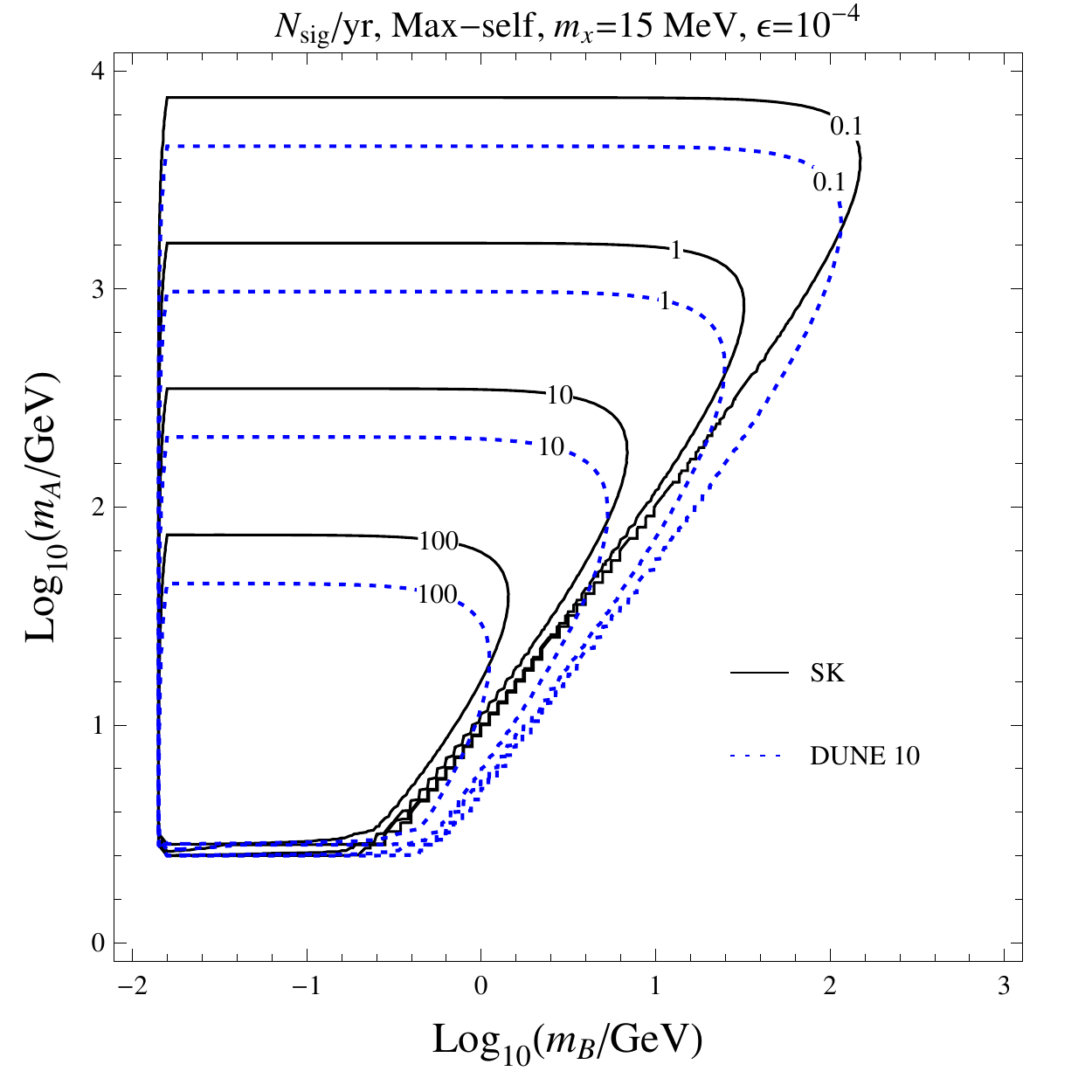}
\end{center}
\vspace*{-.3cm}
\caption{The number of signal events per year at SK and DUNE 10 for $\epsilon^2 = 2 \times10^{-7}$ ($\epsilon = 10^{-4}$) in the top (bottom) panel for Min (Max) SI in the left-panel (right-panel).}
\label{fig:nevents}
\end{figure*}

Using Eq. (\ref{NsigSun}), we calculate the expected number of signal events per year at SK and DUNE 10, which are respectively shown as solid-black and dotted-blue contours in Fig. \ref{fig:nevents} for $\epsilon^2 = 2 \times10^{-7}$ ($\epsilon = 10^{-4}$) in the top (bottom) panel and for Min (Max) SI of $\psi_A$ in the left-panel (right-panel).
Here Min (Max) SI is $\sigma_{AA}/m_A = 0.1 ~(1.25)\,{\rm cm}^2/{\rm g}$.
The interesting shape of the constant number of signal events is well studied in Ref. \cite{Kong:2014mia}.
The boundary in the left side is set by $m_B > m_X$ where $m_X=15$ MeV in our benchmark point ($\log_{10} (m_X/{\rm GeV}) \approx  -1.82$).
The top edge is affected by the DM number density $\propto 1/m_{\rm DM}$.
The right-diagonal edge is determined by $E_{\rm max} > E_{\rm min} = E_{\rm th}$ which is 30 and 100 MeV for DUNE and SK, respectively.
The bottom edge is set by the rapid drop in the accumulated number of DM particles inside the Sun for $m_{\rm DM} \lesssim$ 2--3 GeV due to the active evaporation ($\log_{10} (2.5 ~{\rm GeV}/{\rm GeV}) \approx  0.4$).
The bottom edge is also affected by the energy loss of BDM while traversing the Sun, which is especially active for a larger $\epsilon$. This is shown in the upper panel of Fig. \ref{fig:nevents},
for a smaller value, the effect is weak as illustrated in the bottom panel.

\begin{figure*}[t]
\begin{center}
\hspace*{0.1cm}
\includegraphics[width=0.4\linewidth]{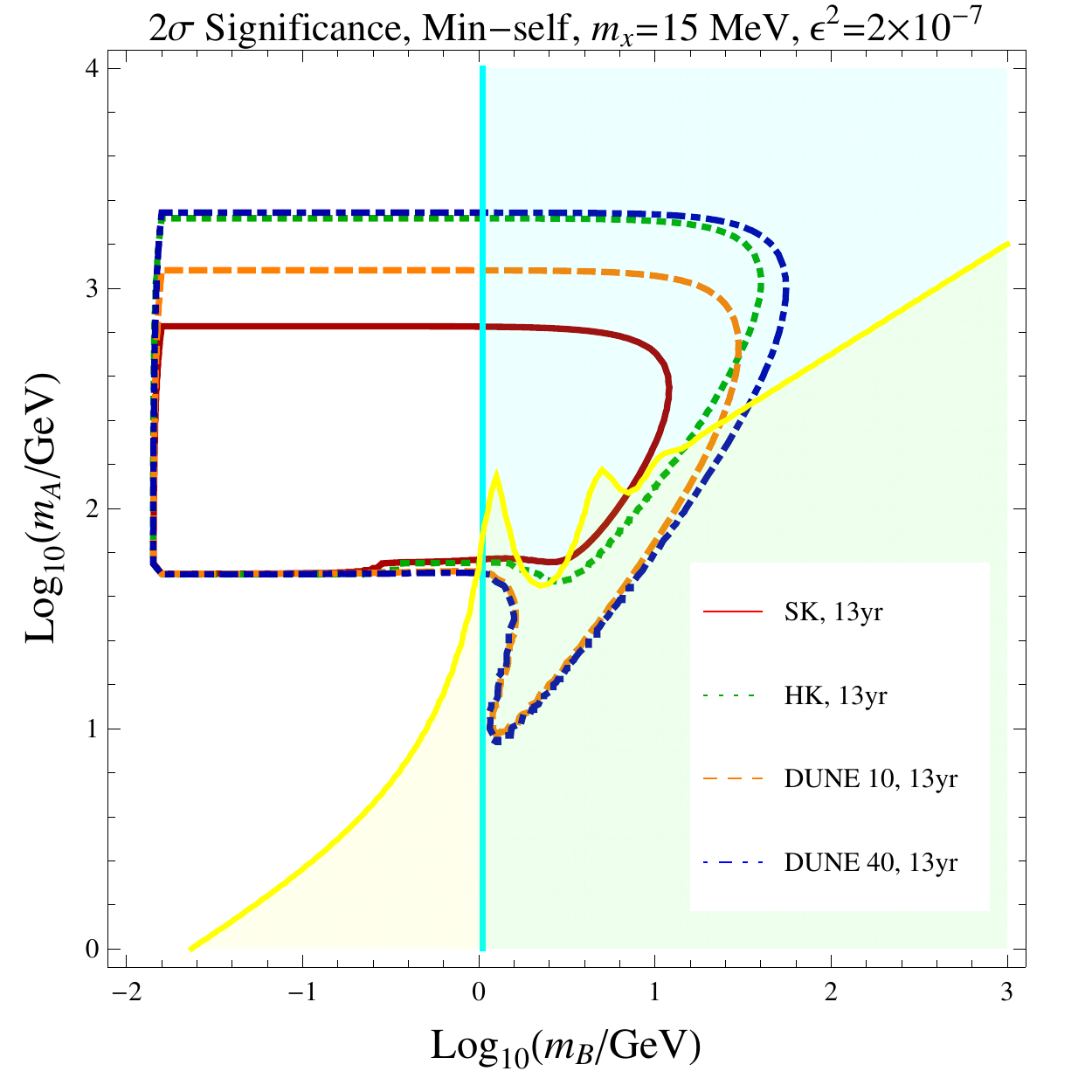}
\hspace*{0.5cm}
\includegraphics[width=0.4\linewidth]{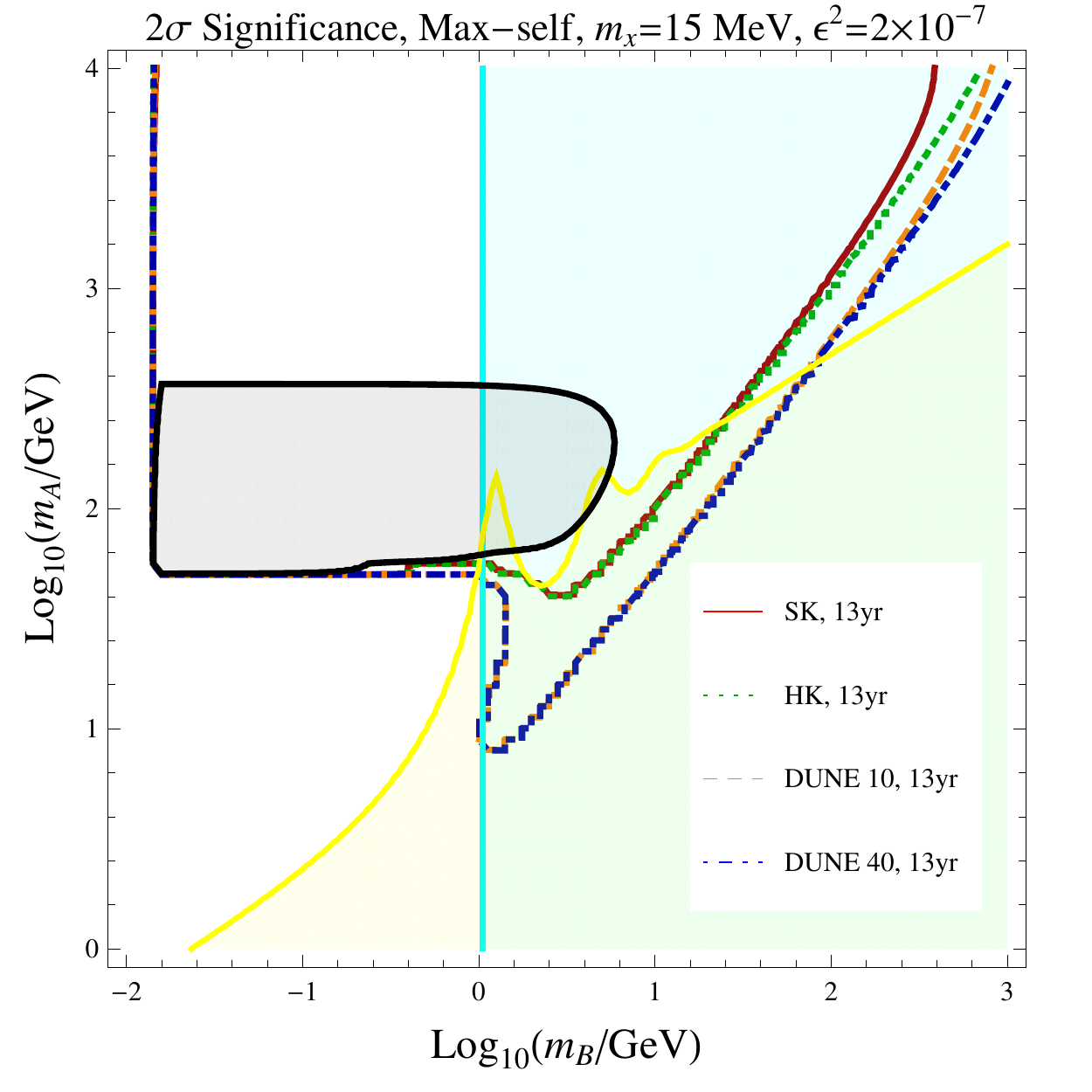} \\
\vspace*{0.2cm}
\includegraphics[width=0.4\linewidth]{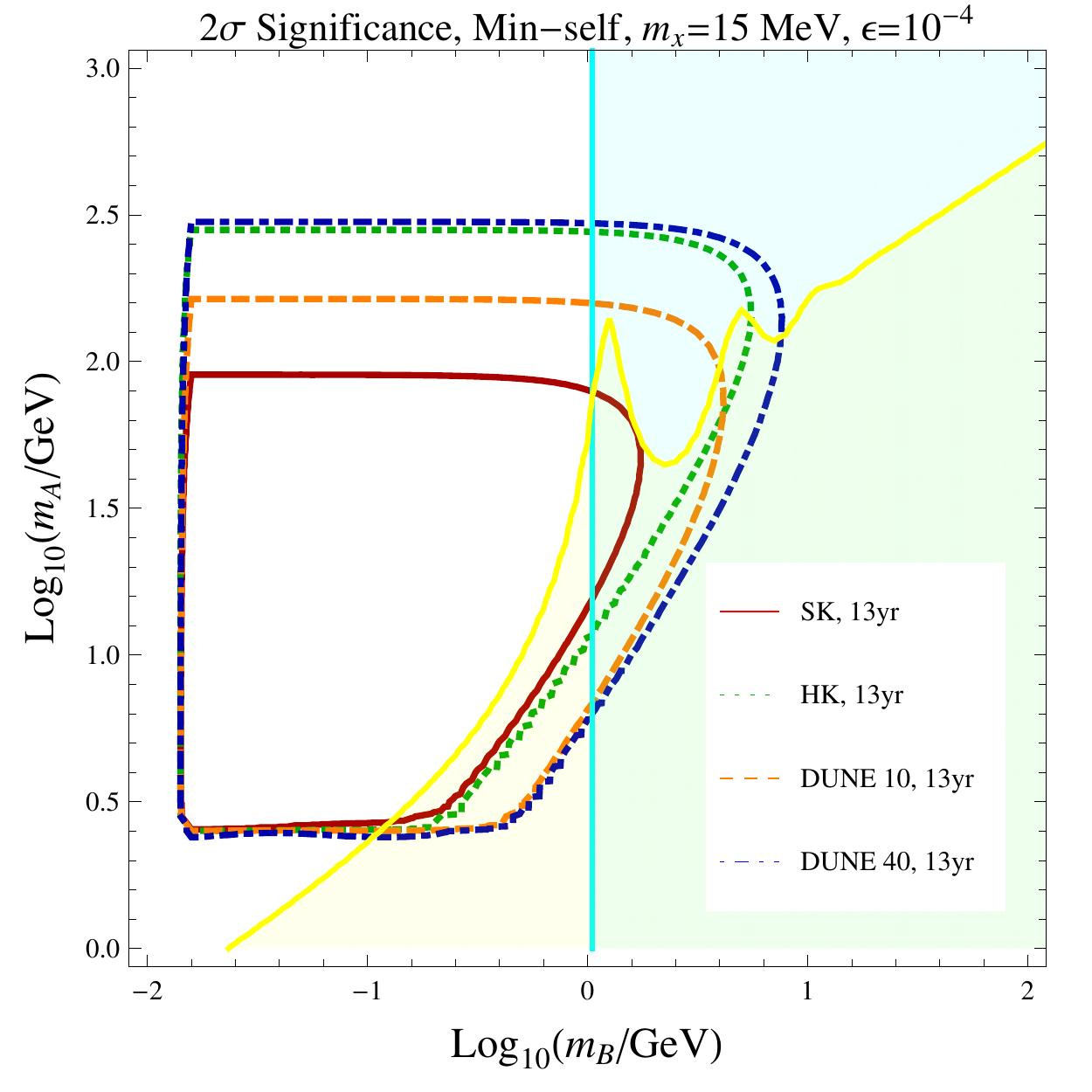}
\hspace*{0.5cm}
\includegraphics[width=0.4\linewidth]{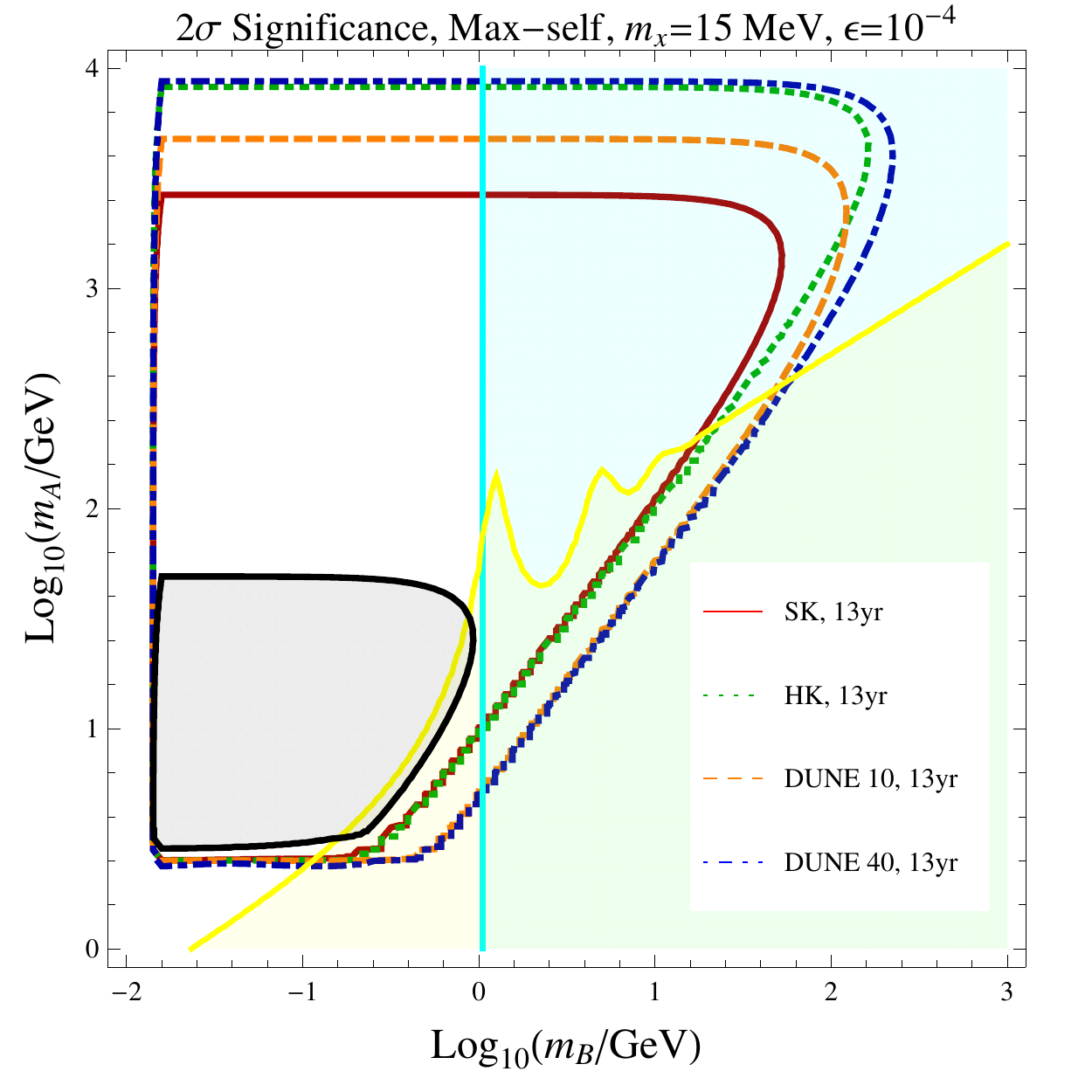}
\end{center}
\vspace*{-.3cm}
\caption{2$\sigma$ significance of various detectors for $\epsilon^2 = 2 \times10^{-7}$ ($\epsilon = 10^{-4}$) in the top (bottom) panel for Min (Max) SI in the left-panel (right-panel). All curves assume 13.6 years of running.}
\label{fig:sun13.6}
\end{figure*}
Using Eq. (\ref{eq:sigma}) and including the background rates given in Table \ref{table:bknd}, we calculate the 2$\sigma$ signal-significance, which is shown in Fig. \ref{fig:sun13.6} for SK, HK and DUNE for $\epsilon^2 = 2 \times10^{-7}$ ($\epsilon = 10^{-4}$) in the top (bottom) panel for Min (Max) SI in the left-panel (right-panel).
We assume that all detectors have been running for 13.6 years.
As shown in Fig. \ref{fig:GC}, we include bounds from the CMB as well as DAMIC.
It turns out that current SK limit applies to maximum self-interaction only.

As shown in the figure, the performance of DUNE 10 is much better than that of SK
and DUNE 10 probes more parameter space: in the upper boundary because of the smaller background due to better angular resolution and along the diagonal direction due to the lower threshold energy.
This implies that the strength of the DUNE detector is more pronounced for Solar BDM than for GC BDM.
Similarly, DUNE 40 is comparable to HK and in fact probes more parameter space along the diagonal direction due to the lower threshold energy.
Again this result illustrates the great performance of the DUNE detector with Solar BDM, even if the volume of the HK detector is about 14 times larger than that of DUNE 40.

For BDM from the GC, the parameter space probed by these detectors is below $m_A \sim 100$ GeV for $\epsilon^2 = 2 \times10^{-7}$ \cite{Necib:2016aez}, while the parameter space even above $m_A \sim 100$ GeV would be covered for BDM arising from the Sun.
See the bottom panels of Fig.~\ref{fig:GC} and the top panels of Fig.~\ref{fig:sun13.6}.

\begin{figure*}[t]
\begin{center}
\hspace*{0.1cm}
\includegraphics[width=0.4\linewidth]{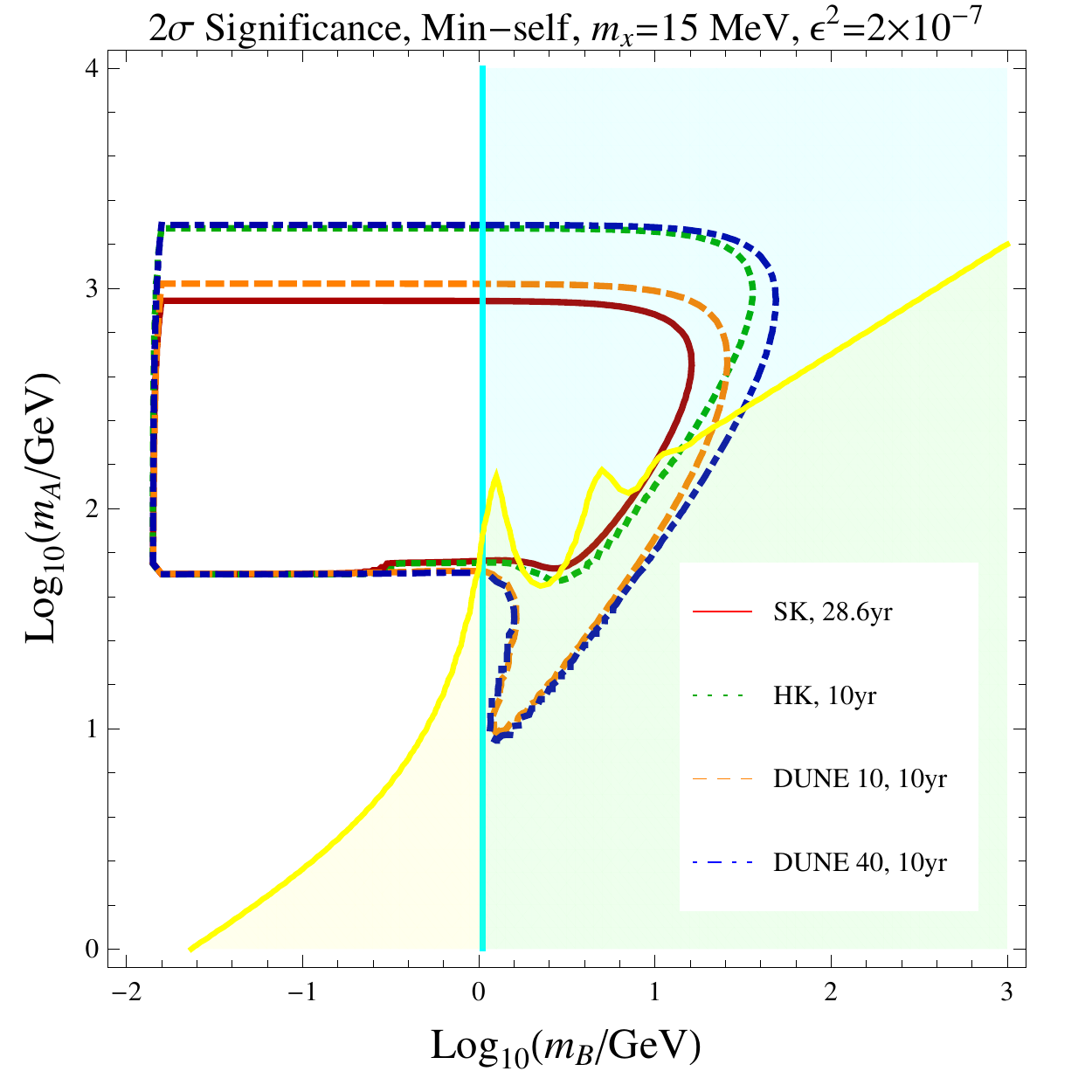}
\hspace*{0.5cm}
\includegraphics[width=0.4\linewidth]{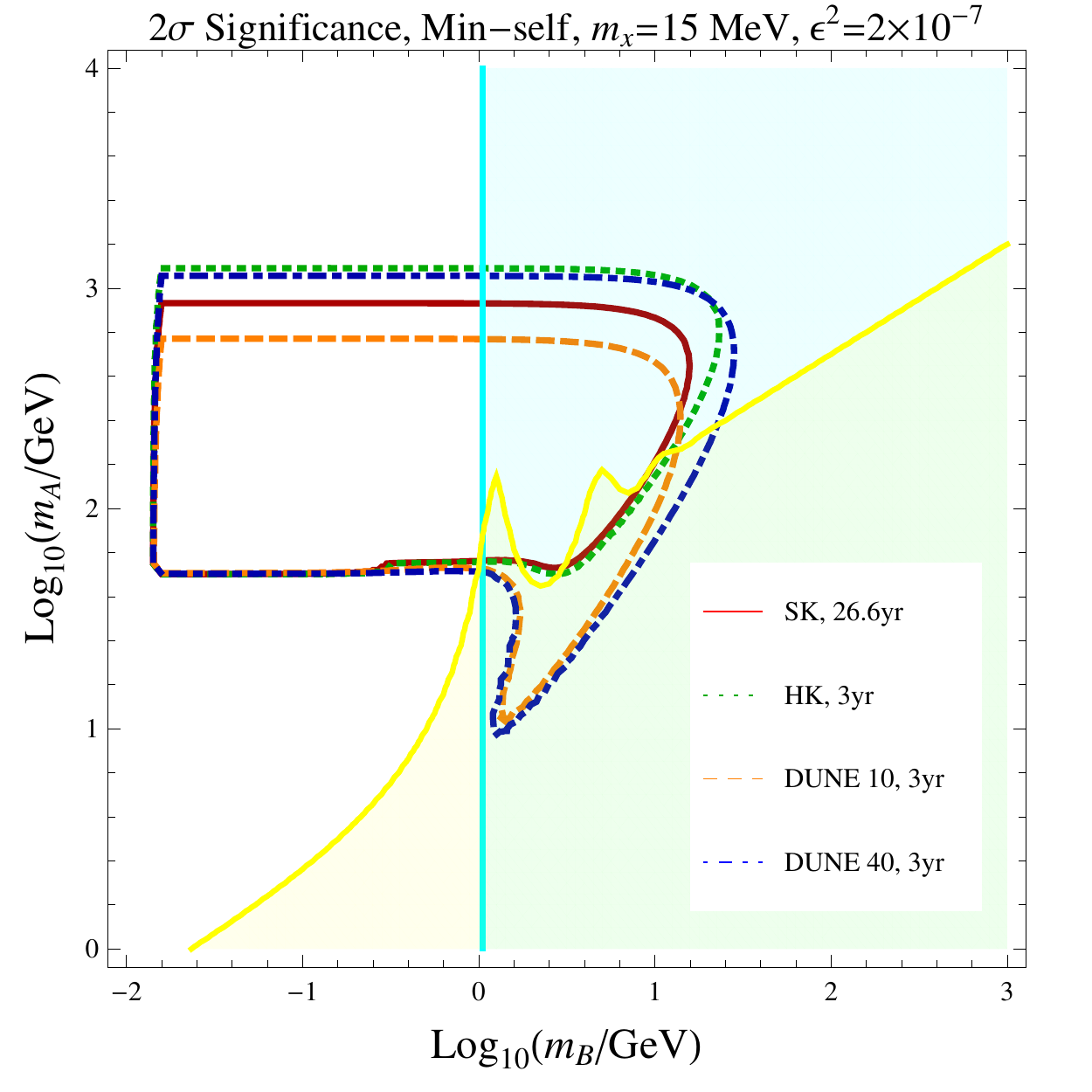} \\
\vspace*{0.2cm}
\includegraphics[width=0.4\linewidth]{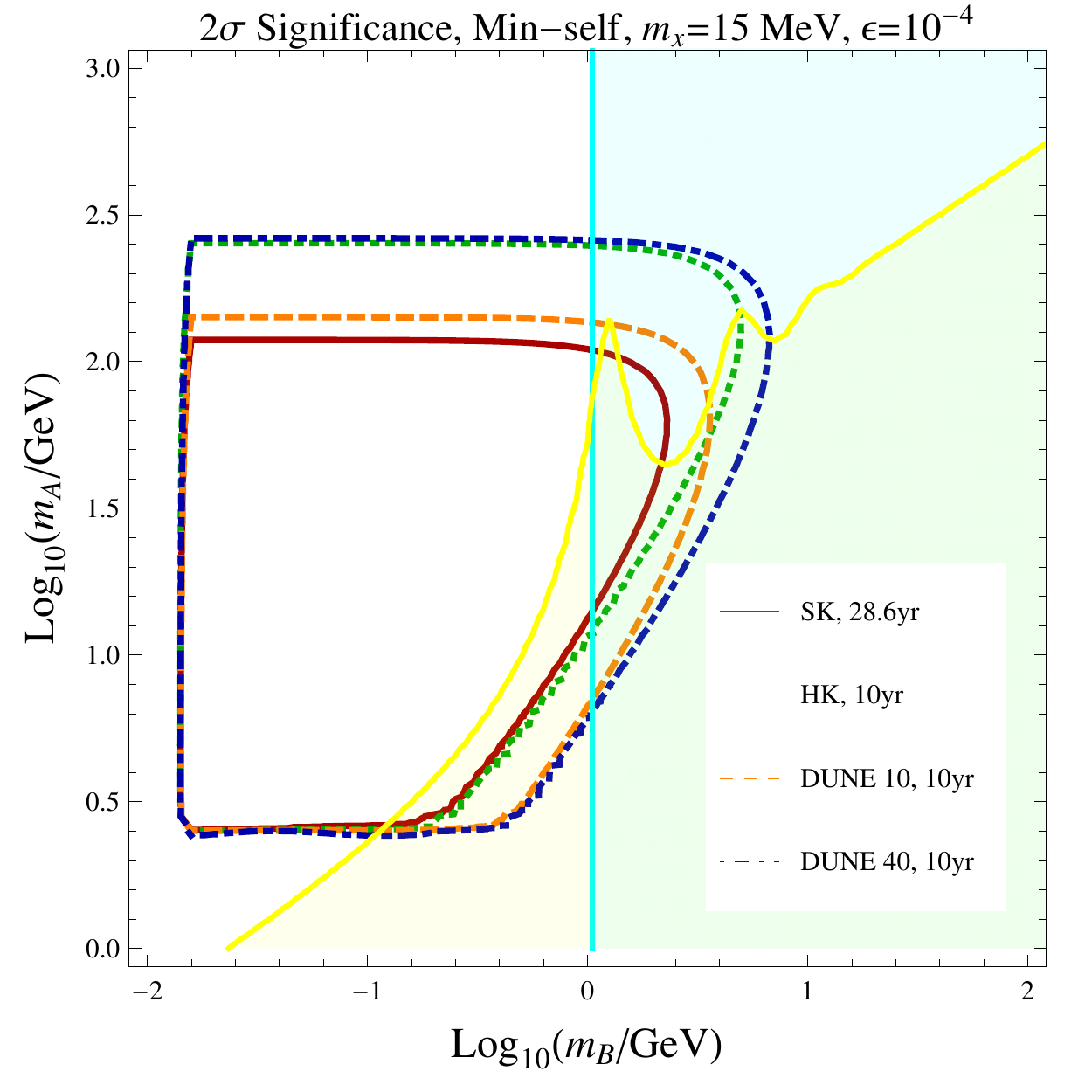}
\hspace*{0.5cm}
\includegraphics[width=0.4\linewidth]{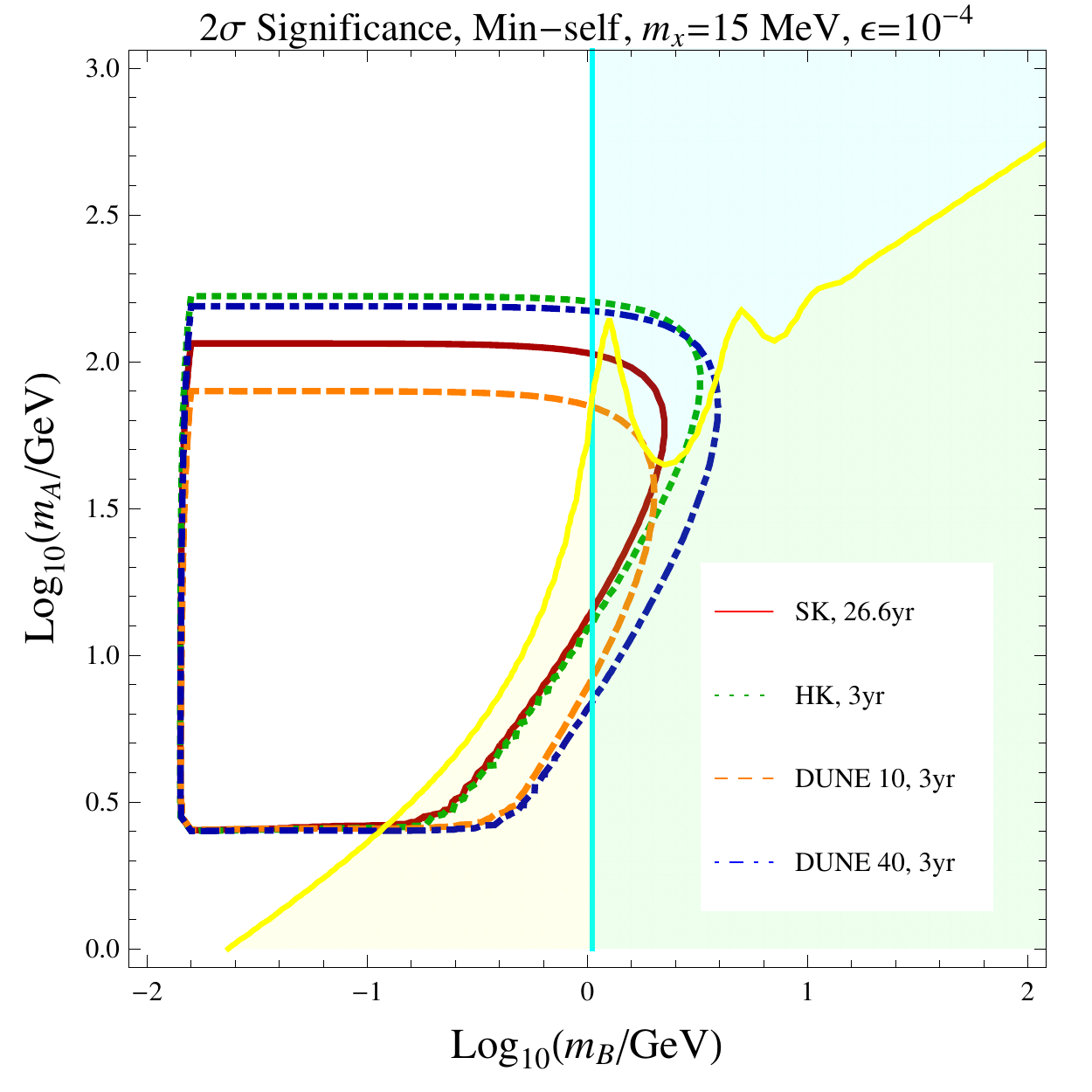}
\end{center}
\vspace*{-.3cm}
\caption{2$\sigma$ significance assuming 5 years of construction and 10 years of physics running of DUNE/HK (left), and 10 years of construction and 3 years of physics running of DUNE/HK (right) for $\epsilon^2 = 2 \times10^{-7}$ ($\epsilon = 10^{-4}$) in the top (bottom) panel. All curves assume Min SI.}
\label{fig:realistic}
\end{figure*}

Fig. \ref{fig:realistic} is the same as Fig. \ref{fig:sun13.6} but for a more realistic timeline.
The 2$\sigma$ significance is shown assuming 5 years of construction and 10 years of physics running of DUNE/HK (left), and 10 years of construction and 3 years of physics running of DUNE/HK (right) for $\epsilon^2 = 2 \times10^{-7}$ ($\epsilon = 10^{-4}$) in the top (bottom) panel.
All curves assume minimum self-interaction, for which the current SK limit is rather weak and does not constrain the $m_A$--$m_B$ space for the given choice of other parameters.
In the second scenario with 10 years of construction and 3 years physics running,
the SK-contour covers a little more in the higher mass (larger $m_A$) due to the longer exposure time, while a slightly larger $m_B$ is probed at DUNE due to a lower threshold energy for a fixed value of $m_A$.
If construction time can be reduced, {\it i.e.,} 5 years, then the signal significance at DUNE is superior as shown in the left panel, even if SK (HK) is (much) larger than DUNE 10 (DUNE 40) in volume.

Comparing results for BDM from the GC as shown in Fig. \ref{fig:GC} (for $\epsilon^2 = 2 \times10^{-7}$), solar BDM is less constrained by the CMB since the relevant parameter space is slightly moved up to a higher $m_A$ region due to the evaporation and the energy loss.
However, a larger portion of the mass space is more constrained by the direct detection of non-relativistic $\psi_B$ ($m_B < 1$ GeV from DAMIC).
On the other hand, BDM from the GC is constrained more by the CMB and less by DAMIC.

\section{Conclusion\label{sec:summary}}

The problem of identifying DM has become central to the fields of particle physics and astrophysics.
While in the coming years DM physics will have a great boost with experimental and technological progresses to put the most promising ideas to the test, we have no firm clue in its identity yet, which naturally leads to a diversity of possible DM candidates.
Among them, especially multi-component DM models are very well-received.
In this paper, we have focused on a scenario where two DM candidates have a large mass gap, with the heavier one as the dominant component in our universe and the lighter one as subdominant.
The heavier candidate is secluded from the SM sector without any tree-level interaction, while the lighter one interacts with the SM via light dark photon exchange. Although subdominant, the lighter DM particles are produced with a large Lorentz boost by the present-day annihilation of the heavier counterpart in the GC or in the center of the Sun.
Its detection prospect at various neutrino telescopes has been examined.
Only very recently, DUNE was considered in terms of BDM coming from the GC.

In our study, we have investigated the discovery potential of DUNE for the BDM arising from the Sun and compared the results with those for large volume neutrino detectors based on Cherenkov radiation such as SK and HK.
LArTPC detectors provide excellent particle identification, which can be used for background reduction in search for a BDM signal.
In particular, a point-like source such as the Sun benefits greatly from the good angular resolution of $1^\circ$, which significantly reduces backgrounds while retaining the same amount of signal events.
A lower threshold energy of 30 MeV also increases the signal sensitivity in the relevant parameter space.
As a result, the strength of the DUNE detector is remarkable, especially for the solar BDM.

Other potential bounds may come from dark photon searches, direct detection of non-relativistic particles (both heavy and light ones), indirect detection of the non-relativistic candidate, CMB constraints on annihilation of the lighter DM candidate, BBN constraints on the lighter one, and DM searches at colliders.
In our scenario with two DM candidates, the most important bounds are obtained from the CMB and direct detection of the lighter DM candidate.
However, these bounds are model-dependent and can be evaded in a different setup.

We have shown that it is very promising to look for BDM particles, especially from the Sun, at DUNE with a LArTPC detector.
We find that the performance of DUNE (10 kTon or 40 kTon) is much better than that of SK or even HK, for the same exposure time, even if their volumes are smaller.
Finally, searches for BDM particles coming from the GC and from the center of the Sun are complementary, since the allowed parameter space that is accessible to one is not to the other.
For instance, $m_A \gtrsim 100$ GeV may be better probed with the solar BDM (from our current study), while
$m_A \lesssim 100$ GeV is well covered for the GC BDM (from Ref. \cite{Necib:2016aez}), for $\epsilon^2 = 2 \times 10^{-7}$.

\section*{Acknowledgements}

We thank Lina Necib for valuable conversations on boosted dark matter and DUNE, and Hai-Bo Yu for providing relevant references on the halo profiles with self-interacting dark matter. We also thank Jeong Han Kim and Brian Batell for helpful discussion.
HA is supported by the scholarship provided by Jazan University in the Kingdom of Saudi Arabia.
KK is supported by US-DOE (DE-FG02-12ER41809) and 
JCP is supported by the National Research Foundation of Korea (NRF-2016R1C1B2015225).
GM is supported in part by the National Research Foundation of South Africa, Grant No. 88614 and by the dissertation fellowship at the University of Kansas.

\bibliography{draft}

\begin{thebibliography}{54}
\expandafter\ifx\csname natexlab\endcsname\relax\def\natexlab#1{#1}\fi
\expandafter\ifx\csname bibnamefont\endcsname\relax
  \def\bibnamefont#1{#1}\fi
\expandafter\ifx\csname bibfnamefont\endcsname\relax
  \def\bibfnamefont#1{#1}\fi
\expandafter\ifx\csname citenamefont\endcsname\relax
  \def\citenamefont#1{#1}\fi
\expandafter\ifx\csname url\endcsname\relax
  \def\url#1{\texttt{#1}}\fi
\expandafter\ifx\csname urlprefix\endcsname\relax\def\urlprefix{URL }\fi
\providecommand{\bibinfo}[2]{#2}
\providecommand{\eprint}[2][]{\url{#2}}

\bibitem[{\citenamefont{Arrenberg et~al.}(2013)}]{Arrenberg:2013rzp}
\bibinfo{author}{\bibfnamefont{S.}~\bibnamefont{Arrenberg}}
  \bibnamefont{et~al.}, in \emph{\bibinfo{booktitle}{{Proceedings, Community
  Summer Study 2013: Snowmass on the Mississippi (CSS2013): Minneapolis, MN,
  USA, July 29-August 6, 2013}}} (\bibinfo{year}{2013}), \eprint{1310.8621},
  \urlprefix\url{https://inspirehep.net/record/1262784/files/arXiv:1310.8621.pdf}.

\bibitem[{\citenamefont{de~Blok}(2010)}]{deBlok:2009sp}
\bibinfo{author}{\bibfnamefont{W.~J.~G.} \bibnamefont{de~Blok}},
  \bibinfo{journal}{Adv. Astron.} \textbf{\bibinfo{volume}{2010}},
  \bibinfo{pages}{789293} (\bibinfo{year}{2010}), \eprint{0910.3538}.

\bibitem[{\citenamefont{Boylan-Kolchin
  et~al.}(2011)\citenamefont{Boylan-Kolchin, Bullock, and
  Kaplinghat}}]{BoylanKolchin:2011de}
\bibinfo{author}{\bibfnamefont{M.}~\bibnamefont{Boylan-Kolchin}},
  \bibinfo{author}{\bibfnamefont{J.~S.} \bibnamefont{Bullock}},
  \bibnamefont{and}
  \bibinfo{author}{\bibfnamefont{M.}~\bibnamefont{Kaplinghat}},
  \bibinfo{journal}{Mon. Not. Roy. Astron. Soc.}
  \textbf{\bibinfo{volume}{415}}, \bibinfo{pages}{L40} (\bibinfo{year}{2011}),
  \eprint{1103.0007}.

\bibitem[{\citenamefont{Boylan-Kolchin
  et~al.}(2012)\citenamefont{Boylan-Kolchin, Bullock, and
  Kaplinghat}}]{BoylanKolchin:2011dk}
\bibinfo{author}{\bibfnamefont{M.}~\bibnamefont{Boylan-Kolchin}},
  \bibinfo{author}{\bibfnamefont{J.~S.} \bibnamefont{Bullock}},
  \bibnamefont{and}
  \bibinfo{author}{\bibfnamefont{M.}~\bibnamefont{Kaplinghat}},
  \bibinfo{journal}{Mon. Not. Roy. Astron. Soc.}
  \textbf{\bibinfo{volume}{422}}, \bibinfo{pages}{1203} (\bibinfo{year}{2012}),
  \eprint{1111.2048}.

\bibitem[{\citenamefont{Lovell et~al.}(2014)\citenamefont{Lovell, Frenk, Eke,
  Jenkins, Gao, and Theuns}}]{Lovell:2013ola}
\bibinfo{author}{\bibfnamefont{M.~R.} \bibnamefont{Lovell}},
  \bibinfo{author}{\bibfnamefont{C.~S.} \bibnamefont{Frenk}},
  \bibinfo{author}{\bibfnamefont{V.~R.} \bibnamefont{Eke}},
  \bibinfo{author}{\bibfnamefont{A.}~\bibnamefont{Jenkins}},
  \bibinfo{author}{\bibfnamefont{L.}~\bibnamefont{Gao}}, \bibnamefont{and}
  \bibinfo{author}{\bibfnamefont{T.}~\bibnamefont{Theuns}},
  \bibinfo{journal}{Mon. Not. Roy. Astron. Soc.}
  \textbf{\bibinfo{volume}{439}}, \bibinfo{pages}{300} (\bibinfo{year}{2014}),
  \eprint{1308.1399}.

\bibitem[{\citenamefont{Spergel and Steinhardt}(2000)}]{Spergel:1999mh}
\bibinfo{author}{\bibfnamefont{D.~N.} \bibnamefont{Spergel}} \bibnamefont{and}
  \bibinfo{author}{\bibfnamefont{P.~J.} \bibnamefont{Steinhardt}},
  \bibinfo{journal}{Phys. Rev. Lett.} \textbf{\bibinfo{volume}{84}},
  \bibinfo{pages}{3760} (\bibinfo{year}{2000}), \eprint{astro-ph/9909386}.

\bibitem[{\citenamefont{Rocha et~al.}(2013)\citenamefont{Rocha, Peter, Bullock,
  Kaplinghat, Garrison-Kimmel, Onorbe, and Moustakas}}]{Rocha:2012jg}
\bibinfo{author}{\bibfnamefont{M.}~\bibnamefont{Rocha}},
  \bibinfo{author}{\bibfnamefont{A.~H.~G.} \bibnamefont{Peter}},
  \bibinfo{author}{\bibfnamefont{J.~S.} \bibnamefont{Bullock}},
  \bibinfo{author}{\bibfnamefont{M.}~\bibnamefont{Kaplinghat}},
  \bibinfo{author}{\bibfnamefont{S.}~\bibnamefont{Garrison-Kimmel}},
  \bibinfo{author}{\bibfnamefont{J.}~\bibnamefont{Onorbe}}, \bibnamefont{and}
  \bibinfo{author}{\bibfnamefont{L.~A.} \bibnamefont{Moustakas}},
  \bibinfo{journal}{Mon. Not. Roy. Astron. Soc.}
  \textbf{\bibinfo{volume}{430}}, \bibinfo{pages}{81} (\bibinfo{year}{2013}),
  \eprint{1208.3025}.

\bibitem[{\citenamefont{Peter et~al.}(2013)\citenamefont{Peter, Rocha, Bullock,
  and Kaplinghat}}]{Peter:2012jh}
\bibinfo{author}{\bibfnamefont{A.~H.~G.} \bibnamefont{Peter}},
  \bibinfo{author}{\bibfnamefont{M.}~\bibnamefont{Rocha}},
  \bibinfo{author}{\bibfnamefont{J.~S.} \bibnamefont{Bullock}},
  \bibnamefont{and}
  \bibinfo{author}{\bibfnamefont{M.}~\bibnamefont{Kaplinghat}},
  \bibinfo{journal}{Mon. Not. Roy. Astron. Soc.}
  \textbf{\bibinfo{volume}{430}}, \bibinfo{pages}{105} (\bibinfo{year}{2013}),
  \eprint{1208.3026}.

\bibitem[{\citenamefont{Randall et~al.}(2008)\citenamefont{Randall, Markevitch,
  Clowe, Gonzalez, and Bradac}}]{Randall:2007ph}
\bibinfo{author}{\bibfnamefont{S.~W.} \bibnamefont{Randall}},
  \bibinfo{author}{\bibfnamefont{M.}~\bibnamefont{Markevitch}},
  \bibinfo{author}{\bibfnamefont{D.}~\bibnamefont{Clowe}},
  \bibinfo{author}{\bibfnamefont{A.~H.} \bibnamefont{Gonzalez}},
  \bibnamefont{and} \bibinfo{author}{\bibfnamefont{M.}~\bibnamefont{Bradac}},
  \bibinfo{journal}{Astrophys. J.} \textbf{\bibinfo{volume}{679}},
  \bibinfo{pages}{1173} (\bibinfo{year}{2008}), \eprint{0704.0261}.

\bibitem[{\citenamefont{Zavala et~al.}(2013)\citenamefont{Zavala, Vogelsberger,
  and Walker}}]{Zavala:2012us}
\bibinfo{author}{\bibfnamefont{J.}~\bibnamefont{Zavala}},
  \bibinfo{author}{\bibfnamefont{M.}~\bibnamefont{Vogelsberger}},
  \bibnamefont{and} \bibinfo{author}{\bibfnamefont{M.~G.}
  \bibnamefont{Walker}}, \bibinfo{journal}{Monthly Notices of the Royal
  Astronomical Society: Letters} \textbf{\bibinfo{volume}{431}},
  \bibinfo{pages}{L20} (\bibinfo{year}{2013}), \eprint{1211.6426}.

\bibitem[{\citenamefont{D'Eramo and Thaler}(2010)}]{D'Eramo:2010ep}
\bibinfo{author}{\bibfnamefont{F.}~\bibnamefont{D'Eramo}} \bibnamefont{and}
  \bibinfo{author}{\bibfnamefont{J.}~\bibnamefont{Thaler}},
  \bibinfo{journal}{JHEP} \textbf{\bibinfo{volume}{06}}, \bibinfo{pages}{109}
  (\bibinfo{year}{2010}), \eprint{1003.5912}.

\bibitem[{\citenamefont{Belanger and Park}(2012)}]{Belanger:2011ww}
\bibinfo{author}{\bibfnamefont{G.}~\bibnamefont{Belanger}} \bibnamefont{and}
  \bibinfo{author}{\bibfnamefont{J.-C.} \bibnamefont{Park}},
  \bibinfo{journal}{JCAP} \textbf{\bibinfo{volume}{1203}}, \bibinfo{pages}{038}
  (\bibinfo{year}{2012}), \eprint{1112.4491}.

\bibitem[{\citenamefont{Belanger et~al.}(2012)\citenamefont{Belanger, Kannike,
  Pukhov, and Raidal}}]{Belanger:2012vp}
\bibinfo{author}{\bibfnamefont{G.}~\bibnamefont{Belanger}},
  \bibinfo{author}{\bibfnamefont{K.}~\bibnamefont{Kannike}},
  \bibinfo{author}{\bibfnamefont{A.}~\bibnamefont{Pukhov}}, \bibnamefont{and}
  \bibinfo{author}{\bibfnamefont{M.}~\bibnamefont{Raidal}},
  \bibinfo{journal}{JCAP} \textbf{\bibinfo{volume}{1204}}, \bibinfo{pages}{010}
  (\bibinfo{year}{2012}), \eprint{1202.2962}.

\bibitem[{\citenamefont{DiFranzo and Mohlabeng}(2016)}]{DiFranzo:2016uzc}
\bibinfo{author}{\bibfnamefont{A.}~\bibnamefont{DiFranzo}} \bibnamefont{and}
  \bibinfo{author}{\bibfnamefont{G.}~\bibnamefont{Mohlabeng}}
  (\bibinfo{year}{2016}), \eprint{1610.07606}.

\bibitem[{\citenamefont{Huang and Zhao}(2014)}]{Huang:2013xfa}
\bibinfo{author}{\bibfnamefont{J.}~\bibnamefont{Huang}} \bibnamefont{and}
  \bibinfo{author}{\bibfnamefont{Y.}~\bibnamefont{Zhao}},
  \bibinfo{journal}{JHEP} \textbf{\bibinfo{volume}{02}}, \bibinfo{pages}{077}
  (\bibinfo{year}{2014}), \eprint{1312.0011}.

\bibitem[{\citenamefont{Agashe et~al.}(2014)\citenamefont{Agashe, Cui, Necib,
  and Thaler}}]{Agashe:2014yua}
\bibinfo{author}{\bibfnamefont{K.}~\bibnamefont{Agashe}},
  \bibinfo{author}{\bibfnamefont{Y.}~\bibnamefont{Cui}},
  \bibinfo{author}{\bibfnamefont{L.}~\bibnamefont{Necib}}, \bibnamefont{and}
  \bibinfo{author}{\bibfnamefont{J.}~\bibnamefont{Thaler}},
  \bibinfo{journal}{JCAP} \textbf{\bibinfo{volume}{1410}}, \bibinfo{pages}{062}
  (\bibinfo{year}{2014}), \eprint{1405.7370}.

\bibitem[{\citenamefont{Berger et~al.}(2015)\citenamefont{Berger, Cui, and
  Zhao}}]{Berger:2014sqa}
\bibinfo{author}{\bibfnamefont{J.}~\bibnamefont{Berger}},
  \bibinfo{author}{\bibfnamefont{Y.}~\bibnamefont{Cui}}, \bibnamefont{and}
  \bibinfo{author}{\bibfnamefont{Y.}~\bibnamefont{Zhao}},
  \bibinfo{journal}{JCAP} \textbf{\bibinfo{volume}{1502}}, \bibinfo{pages}{005}
  (\bibinfo{year}{2015}), \eprint{1410.2246}.

\bibitem[{\citenamefont{Kong et~al.}(2015)\citenamefont{Kong, Mohlabeng, and
  Park}}]{Kong:2014mia}
\bibinfo{author}{\bibfnamefont{K.}~\bibnamefont{Kong}},
  \bibinfo{author}{\bibfnamefont{G.}~\bibnamefont{Mohlabeng}},
  \bibnamefont{and} \bibinfo{author}{\bibfnamefont{J.-C.} \bibnamefont{Park}},
  \bibinfo{journal}{Phys. Lett.} \textbf{\bibinfo{volume}{B743}},
  \bibinfo{pages}{256} (\bibinfo{year}{2015}), \eprint{1411.6632}.

\bibitem[{\citenamefont{Kopp et~al.}(2015)\citenamefont{Kopp, Liu, and
  Wang}}]{Kopp:2015bfa}
\bibinfo{author}{\bibfnamefont{J.}~\bibnamefont{Kopp}},
  \bibinfo{author}{\bibfnamefont{J.}~\bibnamefont{Liu}}, \bibnamefont{and}
  \bibinfo{author}{\bibfnamefont{X.-P.} \bibnamefont{Wang}},
  \bibinfo{journal}{JHEP} \textbf{\bibinfo{volume}{04}}, \bibinfo{pages}{105}
  (\bibinfo{year}{2015}), \eprint{1503.02669}.

\bibitem[{\citenamefont{Necib et~al.}(2016)\citenamefont{Necib, Moon,
  Wongjirad, and Conrad}}]{Necib:2016aez}
\bibinfo{author}{\bibfnamefont{L.}~\bibnamefont{Necib}},
  \bibinfo{author}{\bibfnamefont{J.}~\bibnamefont{Moon}},
  \bibinfo{author}{\bibfnamefont{T.}~\bibnamefont{Wongjirad}},
  \bibnamefont{and} \bibinfo{author}{\bibfnamefont{J.~M.} \bibnamefont{Conrad}}
  (\bibinfo{year}{2016}), \eprint{1610.03486}.

\bibitem[{\citenamefont{Acciarri et~al.}(2015)}]{Acciarri:2015uup}
\bibinfo{author}{\bibfnamefont{R.}~\bibnamefont{Acciarri}} \bibnamefont{et~al.}
  (\bibinfo{collaboration}{DUNE}) (\bibinfo{year}{2015}), \eprint{1512.06148}.

\bibitem[{\citenamefont{Acciarri
  et~al.}(2016{\natexlab{a}})}]{Acciarri:2016ooe}
\bibinfo{author}{\bibfnamefont{R.}~\bibnamefont{Acciarri}} \bibnamefont{et~al.}
  (\bibinfo{collaboration}{DUNE}) (\bibinfo{year}{2016}{\natexlab{a}}),
  \eprint{1601.02984}.

\bibitem[{\citenamefont{Acciarri
  et~al.}(2016{\natexlab{b}})}]{Acciarri:2016crz}
\bibinfo{author}{\bibfnamefont{R.}~\bibnamefont{Acciarri}} \bibnamefont{et~al.}
  (\bibinfo{collaboration}{DUNE}) (\bibinfo{year}{2016}{\natexlab{b}}),
  \eprint{1601.05471}.

\bibitem[{\citenamefont{Strait et~al.}(2016)}]{Strait:2016mof}
\bibinfo{author}{\bibfnamefont{J.}~\bibnamefont{Strait}} \bibnamefont{et~al.}
  (\bibinfo{collaboration}{DUNE}) (\bibinfo{year}{2016}), \eprint{1601.05823}.

\bibitem[{\citenamefont{Okun}(1982)}]{Okun:1982xi}
\bibinfo{author}{\bibfnamefont{L.~B.} \bibnamefont{Okun}},
  \bibinfo{journal}{Sov. Phys. JETP} \textbf{\bibinfo{volume}{56}},
  \bibinfo{pages}{502} (\bibinfo{year}{1982}), \bibinfo{note}{[Zh. Eksp. Teor.
  Fiz.83,892(1982)]}.

\bibitem[{\citenamefont{Holdom}(1986)}]{Holdom:1985ag}
\bibinfo{author}{\bibfnamefont{B.}~\bibnamefont{Holdom}},
  \bibinfo{journal}{Phys. Lett.} \textbf{\bibinfo{volume}{B166}},
  \bibinfo{pages}{196} (\bibinfo{year}{1986}).

\bibitem[{\citenamefont{Huh et~al.}(2008)\citenamefont{Huh, Kim, Park, and
  Park}}]{Huh:2007zw}
\bibinfo{author}{\bibfnamefont{J.-H.} \bibnamefont{Huh}},
  \bibinfo{author}{\bibfnamefont{J.~E.} \bibnamefont{Kim}},
  \bibinfo{author}{\bibfnamefont{J.-C.} \bibnamefont{Park}}, \bibnamefont{and}
  \bibinfo{author}{\bibfnamefont{S.~C.} \bibnamefont{Park}},
  \bibinfo{journal}{Phys. Rev.} \textbf{\bibinfo{volume}{D77}},
  \bibinfo{pages}{123503} (\bibinfo{year}{2008}), \eprint{0711.3528}.

\bibitem[{\citenamefont{Chun and Park}(2009)}]{Chun:2008by}
\bibinfo{author}{\bibfnamefont{E.~J.} \bibnamefont{Chun}} \bibnamefont{and}
  \bibinfo{author}{\bibfnamefont{J.-C.} \bibnamefont{Park}},
  \bibinfo{journal}{JCAP} \textbf{\bibinfo{volume}{0902}}, \bibinfo{pages}{026}
  (\bibinfo{year}{2009}), \eprint{0812.0308}.

\bibitem[{\citenamefont{Chun et~al.}(2011)\citenamefont{Chun, Park, and
  Scopel}}]{Chun:2010ve}
\bibinfo{author}{\bibfnamefont{E.~J.} \bibnamefont{Chun}},
  \bibinfo{author}{\bibfnamefont{J.-C.} \bibnamefont{Park}}, \bibnamefont{and}
  \bibinfo{author}{\bibfnamefont{S.}~\bibnamefont{Scopel}},
  \bibinfo{journal}{JHEP} \textbf{\bibinfo{volume}{02}}, \bibinfo{pages}{100}
  (\bibinfo{year}{2011}), \eprint{1011.3300}.

\bibitem[{\citenamefont{Park and Park}(2013)}]{Park:2012xq}
\bibinfo{author}{\bibfnamefont{J.-C.} \bibnamefont{Park}} \bibnamefont{and}
  \bibinfo{author}{\bibfnamefont{S.~C.} \bibnamefont{Park}},
  \bibinfo{journal}{Phys. Lett.} \textbf{\bibinfo{volume}{B718}},
  \bibinfo{pages}{1401} (\bibinfo{year}{2013}), \eprint{1207.4981}.

\bibitem[{\citenamefont{Belanger et~al.}(2014)\citenamefont{Belanger, Goudelis,
  Park, and Pukhov}}]{Belanger:2013tla}
\bibinfo{author}{\bibfnamefont{G.}~\bibnamefont{Belanger}},
  \bibinfo{author}{\bibfnamefont{A.}~\bibnamefont{Goudelis}},
  \bibinfo{author}{\bibfnamefont{J.-C.} \bibnamefont{Park}}, \bibnamefont{and}
  \bibinfo{author}{\bibfnamefont{A.}~\bibnamefont{Pukhov}},
  \bibinfo{journal}{JCAP} \textbf{\bibinfo{volume}{1402}}, \bibinfo{pages}{020}
  (\bibinfo{year}{2014}), \eprint{1311.0022}.

\bibitem[{\citenamefont{Batley et~al.}(2015)}]{Batley:2015lha}
\bibinfo{author}{\bibfnamefont{J.~R.} \bibnamefont{Batley}}
  \bibnamefont{et~al.} (\bibinfo{collaboration}{NA48/2}),
  \bibinfo{journal}{Phys. Lett.} \textbf{\bibinfo{volume}{B746}},
  \bibinfo{pages}{178} (\bibinfo{year}{2015}), \eprint{1504.00607}.

\bibitem[{\citenamefont{Ilten et~al.}(2015)\citenamefont{Ilten, Thaler,
  Williams, and Xue}}]{Ilten:2015hya}
\bibinfo{author}{\bibfnamefont{P.}~\bibnamefont{Ilten}},
  \bibinfo{author}{\bibfnamefont{J.}~\bibnamefont{Thaler}},
  \bibinfo{author}{\bibfnamefont{M.}~\bibnamefont{Williams}}, \bibnamefont{and}
  \bibinfo{author}{\bibfnamefont{W.}~\bibnamefont{Xue}},
  \bibinfo{journal}{Phys. Rev.} \textbf{\bibinfo{volume}{D92}},
  \bibinfo{pages}{115017} (\bibinfo{year}{2015}), \eprint{1509.06765}.

\bibitem[{\citenamefont{Anastasi et~al.}(2016)}]{Anastasi::2016lwm}
\bibinfo{author}{\bibfnamefont{A.}~\bibnamefont{Anastasi}} \bibnamefont{et~al.}
  (\bibinfo{collaboration}{KLOE-2}), \bibinfo{journal}{Phys. Lett.}
  \textbf{\bibinfo{volume}{B757}}, \bibinfo{pages}{356} (\bibinfo{year}{2016}),
  \eprint{1603.06086}.

\bibitem[{\citenamefont{Banerjee et~al.}(2016)}]{Banerjee:2016tad}
\bibinfo{author}{\bibfnamefont{D.}~\bibnamefont{Banerjee}} \bibnamefont{et~al.}
  (\bibinfo{collaboration}{NA64}) (\bibinfo{year}{2016}), \eprint{1610.02988}.

\bibitem[{\citenamefont{Fechner et~al.}(2009)}]{Fechner:2009aa}
\bibinfo{author}{\bibfnamefont{M.}~\bibnamefont{Fechner}} \bibnamefont{et~al.}
  (\bibinfo{collaboration}{Super-Kamiokande}), \bibinfo{journal}{Phys. Rev.}
  \textbf{\bibinfo{volume}{D79}}, \bibinfo{pages}{112010}
  (\bibinfo{year}{2009}), \eprint{0901.1645}.

\bibitem[{\citenamefont{Kearns et~al.}(2013)}]{Kearns:2013lea}
\bibinfo{author}{\bibfnamefont{E.}~\bibnamefont{Kearns}} \bibnamefont{et~al.}
  (\bibinfo{collaboration}{Hyper-Kamiokande Working Group}), in
  \emph{\bibinfo{booktitle}{{Proceedings, Community Summer Study 2013: Snowmass
  on the Mississippi (CSS2013): Minneapolis, MN, USA, July 29-August 6, 2013}}}
  (\bibinfo{year}{2013}), \eprint{1309.0184},
  \urlprefix\url{https://inspirehep.net/record/1252067/files/arXiv:1309.0184.pdf}.

\bibitem[{\citenamefont{Gaisser and Honda}(2002)}]{Gaisser:2002jj}
\bibinfo{author}{\bibfnamefont{T.~K.} \bibnamefont{Gaisser}} \bibnamefont{and}
  \bibinfo{author}{\bibfnamefont{M.}~\bibnamefont{Honda}},
  \bibinfo{journal}{Ann. Rev. Nucl. Part. Sci.} \textbf{\bibinfo{volume}{52}},
  \bibinfo{pages}{153} (\bibinfo{year}{2002}), \eprint{hep-ph/0203272}.

\bibitem[{\citenamefont{Bays et~al.}(2012)}]{Bays:2011si}
\bibinfo{author}{\bibfnamefont{K.}~\bibnamefont{Bays}} \bibnamefont{et~al.}
  (\bibinfo{collaboration}{Super-Kamiokande}), \bibinfo{journal}{Phys. Rev.}
  \textbf{\bibinfo{volume}{D85}}, \bibinfo{pages}{052007}
  (\bibinfo{year}{2012}), \eprint{1111.5031}.

\bibitem[{\citenamefont{Abe et~al.}(2011)}]{Abe:2010hy}
\bibinfo{author}{\bibfnamefont{K.}~\bibnamefont{Abe}} \bibnamefont{et~al.}
  (\bibinfo{collaboration}{Super-Kamiokande}), \bibinfo{journal}{Phys. Rev.}
  \textbf{\bibinfo{volume}{D83}}, \bibinfo{pages}{052010}
  (\bibinfo{year}{2011}), \eprint{1010.0118}.

\bibitem[{\citenamefont{Pik}(2012)}]{Pik:2012qsy}
\bibinfo{author}{\bibfnamefont{L.~K.} \bibnamefont{Pik}}, Ph.D. thesis,
  \bibinfo{school}{Tokyo U.} (\bibinfo{year}{2012}),
  \urlprefix\url{http://www-sk.icrr.u-tokyo.ac.jp/sk/pub/index.html#dthesis}.

\bibitem[{\citenamefont{Dziomba}(2012)}]{Dziomba:2012paz}
\bibinfo{author}{\bibfnamefont{M.~R.} \bibnamefont{Dziomba}}, Ph.D. thesis,
  \bibinfo{school}{Washington U., Seattle} (\bibinfo{year}{2012}),
  \urlprefix\url{http://www-sk.icrr.u-tokyo.ac.jp/sk/pub/index.html#dthesis}.

\bibitem[{\citenamefont{Richard et~al.}(2016)}]{Richard:2015aua}
\bibinfo{author}{\bibfnamefont{E.}~\bibnamefont{Richard}} \bibnamefont{et~al.}
  (\bibinfo{collaboration}{Super-Kamiokande}), \bibinfo{journal}{Phys. Rev.}
  \textbf{\bibinfo{volume}{D94}}, \bibinfo{pages}{052001}
  (\bibinfo{year}{2016}), \eprint{1510.08127}.

\bibitem[{\citenamefont{Abe et~al.}(2016)}]{T2HKK}
\bibinfo{author}{\bibfnamefont{K.}~\bibnamefont{Abe}} \bibnamefont{et~al.}
  (\bibinfo{collaboration}{Hyper-Kamiokande proto}) (\bibinfo{year}{2016}),
  \eprint{1611.06118}.

\bibitem[{\citenamefont{Navarro et~al.}(1996)\citenamefont{Navarro, Frenk, and
  White}}]{Navarro:1995iw}
\bibinfo{author}{\bibfnamefont{J.~F.} \bibnamefont{Navarro}},
  \bibinfo{author}{\bibfnamefont{C.~S.} \bibnamefont{Frenk}}, \bibnamefont{and}
  \bibinfo{author}{\bibfnamefont{S.~D.~M.} \bibnamefont{White}},
  \bibinfo{journal}{Astrophys. J.} \textbf{\bibinfo{volume}{462}},
  \bibinfo{pages}{563} (\bibinfo{year}{1996}), \eprint{astro-ph/9508025}.

\bibitem[{\citenamefont{Navarro et~al.}(1997)\citenamefont{Navarro, Frenk, and
  White}}]{Navarro:1996gj}
\bibinfo{author}{\bibfnamefont{J.~F.} \bibnamefont{Navarro}},
  \bibinfo{author}{\bibfnamefont{C.~S.} \bibnamefont{Frenk}}, \bibnamefont{and}
  \bibinfo{author}{\bibfnamefont{S.~D.~M.} \bibnamefont{White}},
  \bibinfo{journal}{Astrophys. J.} \textbf{\bibinfo{volume}{490}},
  \bibinfo{pages}{493} (\bibinfo{year}{1997}), \eprint{astro-ph/9611107}.

\bibitem[{\citenamefont{Kaplinghat et~al.}(2015)\citenamefont{Kaplinghat,
  Linden, and Yu}}]{Kaplinghat:2015gha}
\bibinfo{author}{\bibfnamefont{M.}~\bibnamefont{Kaplinghat}},
  \bibinfo{author}{\bibfnamefont{T.}~\bibnamefont{Linden}}, \bibnamefont{and}
  \bibinfo{author}{\bibfnamefont{H.-B.} \bibnamefont{Yu}},
  \bibinfo{journal}{Phys. Rev. Lett.} \textbf{\bibinfo{volume}{114}},
  \bibinfo{pages}{211303} (\bibinfo{year}{2015}), \eprint{1501.03507}.

\bibitem[{\citenamefont{Kaplinghat et~al.}(2014)\citenamefont{Kaplinghat,
  Keeley, Linden, and Yu}}]{Kaplinghat:2013xca}
\bibinfo{author}{\bibfnamefont{M.}~\bibnamefont{Kaplinghat}},
  \bibinfo{author}{\bibfnamefont{R.~E.} \bibnamefont{Keeley}},
  \bibinfo{author}{\bibfnamefont{T.}~\bibnamefont{Linden}}, \bibnamefont{and}
  \bibinfo{author}{\bibfnamefont{H.-B.} \bibnamefont{Yu}},
  \bibinfo{journal}{Phys. Rev. Lett.} \textbf{\bibinfo{volume}{113}},
  \bibinfo{pages}{021302} (\bibinfo{year}{2014}), \eprint{1311.6524}.

\bibitem[{\citenamefont{Kaplinghat et~al.}(2016)\citenamefont{Kaplinghat,
  Tulin, and Yu}}]{Kaplinghat:2015aga}
\bibinfo{author}{\bibfnamefont{M.}~\bibnamefont{Kaplinghat}},
  \bibinfo{author}{\bibfnamefont{S.}~\bibnamefont{Tulin}}, \bibnamefont{and}
  \bibinfo{author}{\bibfnamefont{H.-B.} \bibnamefont{Yu}},
  \bibinfo{journal}{Phys. Rev. Lett.} \textbf{\bibinfo{volume}{116}},
  \bibinfo{pages}{041302} (\bibinfo{year}{2016}), \eprint{1508.03339}.

\bibitem[{\citenamefont{Kim et~al.}(2016)\citenamefont{Kim, Kong, Lee, and
  Mohlabeng}}]{Kim:2016plm}
\bibinfo{author}{\bibfnamefont{J.~H.} \bibnamefont{Kim}},
  \bibinfo{author}{\bibfnamefont{K.}~\bibnamefont{Kong}},
  \bibinfo{author}{\bibfnamefont{S.~J.} \bibnamefont{Lee}}, \bibnamefont{and}
  \bibinfo{author}{\bibfnamefont{G.}~\bibnamefont{Mohlabeng}},
  \bibinfo{journal}{Phys. Rev.} \textbf{\bibinfo{volume}{D94}},
  \bibinfo{pages}{035023} (\bibinfo{year}{2016}), \eprint{1604.07421}.

\bibitem[{\citenamefont{de~Mello~Neto et~al.}(2016)}]{deMelloNeto:2015mca}
\bibinfo{author}{\bibfnamefont{J.~R.~T.} \bibnamefont{de~Mello~Neto}}
  \bibnamefont{et~al.} (\bibinfo{collaboration}{DAMIC}), \bibinfo{journal}{PoS}
  \textbf{\bibinfo{volume}{ICRC2015}}, \bibinfo{pages}{1221}
  (\bibinfo{year}{2016}), \eprint{1510.02126}.

\bibitem[{\citenamefont{Essig et~al.}(2012{\natexlab{a}})\citenamefont{Essig,
  Mardon, and Volansky}}]{Essig:2011nj}
\bibinfo{author}{\bibfnamefont{R.}~\bibnamefont{Essig}},
  \bibinfo{author}{\bibfnamefont{J.}~\bibnamefont{Mardon}}, \bibnamefont{and}
  \bibinfo{author}{\bibfnamefont{T.}~\bibnamefont{Volansky}},
  \bibinfo{journal}{Phys. Rev.} \textbf{\bibinfo{volume}{D85}},
  \bibinfo{pages}{076007} (\bibinfo{year}{2012}{\natexlab{a}}),
  \eprint{1108.5383}.

\bibitem[{\citenamefont{Essig et~al.}(2012{\natexlab{b}})\citenamefont{Essig,
  Manalaysay, Mardon, Sorensen, and Volansky}}]{Essig:2012yx}
\bibinfo{author}{\bibfnamefont{R.}~\bibnamefont{Essig}},
  \bibinfo{author}{\bibfnamefont{A.}~\bibnamefont{Manalaysay}},
  \bibinfo{author}{\bibfnamefont{J.}~\bibnamefont{Mardon}},
  \bibinfo{author}{\bibfnamefont{P.}~\bibnamefont{Sorensen}}, \bibnamefont{and}
  \bibinfo{author}{\bibfnamefont{T.}~\bibnamefont{Volansky}},
  \bibinfo{journal}{Phys. Rev. Lett.} \textbf{\bibinfo{volume}{109}},
  \bibinfo{pages}{021301} (\bibinfo{year}{2012}{\natexlab{b}}),
  \eprint{1206.2644}.

\bibitem[{\citenamefont{Chen et~al.}(2014)\citenamefont{Chen, Lee, Lin, and
  Lin}}]{Chen:2014oaa}
\bibinfo{author}{\bibfnamefont{C.-S.} \bibnamefont{Chen}},
  \bibinfo{author}{\bibfnamefont{F.-F.} \bibnamefont{Lee}},
  \bibinfo{author}{\bibfnamefont{G.-L.} \bibnamefont{Lin}}, \bibnamefont{and}
  \bibinfo{author}{\bibfnamefont{Y.-H.} \bibnamefont{Lin}},
  \bibinfo{journal}{JCAP} \textbf{\bibinfo{volume}{1410}}, \bibinfo{pages}{049}
  (\bibinfo{year}{2014}), \eprint{1408.5471}.

\end{thebibliography}

\end{document}